\documentclass[12pt]{article}
\usepackage{cite}
\usepackage{mathptmx}
\usepackage{setspace}

\usepackage{mathrsfs}

\usepackage{changepage}
\usepackage{dcolumn}
\usepackage[table]{xcolor}
\usepackage{floatrow}
\usepackage{floatrow}   
\usepackage{subcaption}
\usepackage{graphicx,xcolor} 
\usepackage{url}
\usepackage{authblk}
\usepackage{hyperref}

\usepackage[margin=1in]{geometry}
\usepackage{amsmath}
\usepackage{amssymb}
\usepackage{pgf}
\usepackage{comment}

\definecolor{mynavy}{HTML}{000080}
\definecolor{darkred}{HTML}{8B0000}
\definecolor{mygreen}{HTML}{006400}
\definecolor{mygold}{HTML}{B8860B}

\newcolumntype{d}[1]{D..{#1}}

\title{Generalised geometric Brownian motion: Theory and applications to option pricing}

\author{Viktor Stojkoski$^{1,2}$, Trifce Sandev$^{2,3,4}$, Lasko Basnarkov$^{2,5}$, Ljupco Kocarev$^{2,5}$ and Ralf Metzler$^{3,}$\footnote{Corresponding author: rmetzler@uni-potsdam.de}}

\affil{%
$^{1}$Faculty of Economics, Ss.~Cyril and Methodius University, 1000 Skopje, Macedonia\\
$^{2}$Research Center for Computer Science and Information Technologies, Macedonian Academy of Sciences and Arts, Bul. Krste Misirkov 2, 1000 Skopje, Macedonia\\ $^{3}$Institute of Physics \& Astronomy, University of Potsdam, D-14776 Potsdam-Golm, Germany\\ $^{4}$Institute of Physics, Faculty of Natural Sciences and Mathematics, Ss.~Cyril and Methodius University, Arhimedova 3, 1000 Skopje, Macedonia\\ $^{5}$Faculty of Computer Science and Engineering, Ss.~Cyril and Methodius University, P.O. Box 393, 1000 Skopje, Macedonia}

\begin{document}
\maketitle
\begin{abstract}
Classical option pricing schemes assume that the value of a financial asset follows a geometric Brownian motion (GBM). However, a growing body of studies suggest that a simple GBM trajectory is not an adequate representation for asset dynamics due to irregularities found when comparing its properties with empirical distributions. 
As a solution, we develop a generalisation of GBM where the introduction of a memory kernel critically determines the behavior of the stochastic process. We find the general expressions for the moments, log-moments, and the expectation of the periodic log returns, and obtain the corresponding probability density functions by using the subordination approach. Particularly, we consider subdiffusive GBM (sGBM), tempered sGBM, a mix of GBM and sGBM, and a mix of sGBMs. We utilise the resulting generalised GBM (gGBM) to examine the empirical performance of a selected group of kernels in the pricing of European call options. Our results indicate that the performance of a kernel ultimately depends on the maturity of the option and its moneyness.
\end{abstract}




\section{Introduction}\label{sec1}

Geometric Brownian motion (GBM) frequently features in mathematical modeling. The advantage of modelling through this process lies in its universality, as it represents an attractor of more complex models that exhibit non-ergodic dynamics~\cite{stojkoski2019cooperation,stojkoski2019evolution,peters2013ergodicity}. As such, GBM has been used to underlie the dynamics of a diverse set of natural phenomena including the distribution of incomes, body weights, rainfall, fragment sizes in rock crushing processes, etc~\cite{aitchison1957lognormal,redner1990random}. Nevertheless, perhaps the best-known application of GBM is in finance, and in particular in terms of the Black-Scholes (BS) model (or Black-Scholes-Merton model) \cite{black1973pricing,merton1975optimum,merton1976option} for the pricing of European options.

By construction, GBM is a simple continuous-time stochastic process in which the logarithm of the randomly varying quantity of interest follows a Brownian motion with drift. Its non-ergodicity is manifested in the difference between the growth rate observed in an individual trajectory and the ensemble average growth~\cite{1}. The time-averaged growth rate is dependent on both the drift and the randomness in the system, whereas the ensemble growth rate is solely dependent on the drift. If only a single system is to be modeled, on the long run only the time-averaged growth rate, is observed. This is naturally the case in financial market dynamics, for which only single time series exist, and where individual realisations would be expected to be distinctly disparate \cite{oshanin2012two}.

Moreover, GBM is closely related to the problem of heterogeneous diffusion and turbulent diffusion, which are represented by the inhomogeneous advection-diffusion equation with position-dependent diffusion coefficient $D(x)$ and velocity field $v(x)$. It is well known that at turbulent diffusion the contaminant spreads very fast. For the case of Richardson diffusion the position-dependent diffusion coefficient behaves as $D(x)\sim x^{4/3}$ and the relative mean squared displacement (MSD) scales as $\langle x^{2}(t)\rangle\sim t^3$ \cite{procaccia}. However, the fast spread of contaminants can be essentially increased due to multiplicative noise, such that the MSD grows exponentially with time~\cite{7,iomin_prl}.

Notably, in a variety of cases GBM has failed to reproduce the properties of real asset prices. For instance, by definition, GBM is not able to adequately reproduce fat tailed distributions of various characteristics widespread in nature~\cite{taleb2007black}. As a solution, three alternating theories have been proposed: i) stochastic volatility~\cite{cox1975notes, heston1993closed,hagan2002managing}; ii) utilising stochastic processes in which the noise follows a fat-tailed distribution~\cite{matacz2000financial,borland2002theory, borland2004non,moriconi2007delta,cassidy2010pricing,basnarkov2018option}; and iii) generalisations of GBM based on subdiffusion~\cite{13,14,krzyzanowski2020weighted}. In the first approach the volatility is a stochastic process itself. The second approach intuitively leads to the observation of log returns which follow a fat-tailed distribution. The last approach, differently from the first two views, assumes anomalous price dynamics. Concretely, the observation that the distribution of log returns is fat tailed, can be attributed to prolonged periods in which the price of the asset exhibits approximately constant extreme values. These constant periods can be considered as trapping of particles, as is done in physical systems which manifest anomalous diffusion (subdiffusion) \cite{scalas2000fractional,raberto2002waiting}. While the resulting {\it subdiffusive} GBM (sGBM) is able to easily reproduce real-life properties, the literature lacks a extensive study in which the exact empirical characteristics of the subdiffusive model are presented.

The purpose of this paper is to propose a unifying framework for the application of subdiffusive GBM models in option pricing. We do this by developing the so-called generalised GBM (gGBM). gGBM is a stochastic process whose behavior is critically determined by a memory kernel. By choosing the appropriate kernel, we recover the standard GBM and the typically used subdiffusive GBM models~\cite{13,14,krzyzanowski2020weighted}. To understand the behavior of gGBM under various kernels, we perform a detailed analysis and show that the dynamics of the model can be easily adjusted to mimic periods of constant prices and/or fat-tailed observations of returns, thus corresponding to realistic scenarios. More importantly, we utilise the properties of the model to investigate its capability to predict empirical option values. We find that the performance of a kernel ultimately depends on the parameters of the option, such as its maturity and its moneyness. The first property describes simply the time left for the option to be exercised, wheres the second characteristic depicts the relative position of the current price with respect to the strike price of the option. On the first sight, this conclusion appears intuitive -- obviously the known information for the properties of the asset greatly impacts its price, the observation that a slight change in the known information may drastically change the dynamics suggests that there is a need in the option pricing literature for models that easily allow for such structural changes. We believe that the resolution to this issue lies in applying the concepts of time-averaging and ergodicity breaking to modeling financial time-series, and our gGBM framework offers a computationally inexpensive and efficiently tractable solution.

The paper is organised as follows. In Section~\ref{sec:background} we provide an overview of GBM in the BS model and its use in option pricing. We also give detailed results for the so-called sGBM in terms of fractional Fokker-Planck equation and its corresponding continuous time random walk (CTRW) model. In Section~\ref{sec:gen-gbm} we present gGBM and describe its properties by using the subordination approach. In particular, we derive the corresponding Fokker-Planck equation with a memory kernel and obtain the respective moments and log-moments. The general function used in the L\'evy exponent occurs as a memory kernel in the Fokker-Planck equation, which allows us to recover the previously known results for GBM and sGBM. We consider generalisations of GBM and sGBM by introducing tempered sGBM, a mix of GBM and sGBM, as well as a mix of sGBMs. An empirical example of application of the gGBM in option pricing is presented in Section~\ref{sec:empirical-example}. Section~\ref{sec:conclusion} summarises our findings. In the Appendices we give detailed calculations as well as derivation of the Fokker-Planck equation for the gGBM within the CTRW theory.


\section{Background}\label{sec:background}

\subsection{Standard GBM}

GBM has been applied in a variety scientific fields \cite{1,2,3,4,5,aw}. Mathematically, it is represented by the Langevin equation
\begin{align}\label{eq:langevin}
dx(t)=x(t)\left[\mu\,dt+\sigma\,dB(t)\right], \quad x_0=x(0),
\end{align}
where $x(t)$ is the particle position, $\mu$ is the drift, $\sigma>0$ is the volatility, and $B(t)$ represents standard Brownian motion. The solution to Eq.~(\ref{eq:langevin}) in the It\^{o} sense is
\begin{align}\label{eq:GBM-solution}
x(t)=x_0\,e^{(\mu - \frac{\sigma^2}{2}) t + \sigma B(t)}, \quad x_0=x(0)>0.
\end{align}

When the dynamics of the asset price follows a GBM, then a risk-neutral distribution (probability distribution which takes into
account the risk of future price fluctuations) can be easily found by solving the corresponding Fokker-Planck equation\footnote{This Fokker-Planck equation corresponds to the It\^{o} interpretation of the multiplicative noise. There are also Stratonovich and Klimontovich-H\"{a}nggi interpretations, for which the corresponding Fokker-Planck equations are slightly different, see Refs.~\cite{6,7}. In finance math literature the It\^{o} convention is the standard interpretation.} to Eq.~(\ref{eq:langevin}),
\begin{align}\label{generalizedFPE_standard}
\frac{\partial}{\partial t}f(x,t)=-\mu\frac{\partial}{\partial x}xf(x,t)+\frac{\sigma^2}{2}\frac{\partial^2}{\partial x^2}x^{2}f(x,t),
\end{align}
with initial condition $f(x,t=0)=\delta(x-x_0)$. The solution of Eq.~(\ref{generalizedFPE_standard}) is the famed log-normal distribution
\begin{align}\label{lognormal distribution}
    f(x,t)=\frac{1}{x\sqrt{2\pi\sigma^2t}}\times\exp\left(-\frac{\left[\log{x}-\log{x_0}-\bar{\mu}\,t\right]^{2}}{2\sigma^{2}t}\right).
\end{align}
where $\bar{\mu}=\mu-\sigma^{2}/2$.

From the solution, it follows that the mean value and the mean square displacement (MSD) have exponential dependence on time,
\begin{align}
    \langle x(t)\rangle=x_0\,e^{\mu\,t},
\end{align}
and
\begin{align}
    \langle x^{2}(t)\rangle=x_{0}^{2}\,e^{(\sigma^{2}+2\,\mu)t},
\end{align}
respectively, and thus, the variance becomes
\begin{align}
    \langle x^{2}(t)\rangle-\left\langle x(t)\right\rangle^{2}= x_{0}^{2}\,e^{2\mu t}\left(e^{\sigma^{2}t}-1\right).
\end{align}
The exact derivation of the GBM distribution and its moments is given in Appendix~\ref{app1}.

Evidently, in GBM the diffusion coefficient scales proportionally with the square of the position of the particle, i.e., $D(x)=\sigma^2 x^2/2$, and thus the MSD has an exponential dependence on time. A more convenient measure instead of the MSD for geometric processes is the behaviour of the expectation of the logarithm of $x(t)$. In the case of GBM the expectation of the logarithm of the particle position has a linear dependence on time. This can be shown by calculation of the log-moments $\left\langle \log^{n}{x(t)}\right\rangle=\int_{0}^{\infty}\log^{n}x\,P(x,t)\,dx$, see Appendix~\ref{app1}. The mean value of the logarithm of $x(t)$ becomes
\begin{align}
    \langle \log{x(t)}\rangle=\langle \log{x_0}\rangle+\bar{\mu}\,t,
\end{align}
from where for the expectation of the periodic log return with period $\Delta{t}$, one finds
\begin{align}
    \frac{1}{\Delta{t}}\langle \log{\left(x(t+\Delta{t})/x(t)\right)}\rangle\underset{\Delta t\rightarrow0}{\sim}\bar{\mu}=\frac{d}{dt}\langle\log{x(t)}\rangle.
\end{align}
The second log-moment is given by
\begin{align}
    \langle \log^{2}{x(t)}\rangle=\langle \log^{2}{x_0}\rangle+\bar{\mu}^{2}t^{2}+2\bar{\mu}\langle \log{x_0}\rangle t+\sigma^{2}t,
\end{align}
which for the log-variance yields
\begin{align}
    \langle \log^{2}{x(t)}\rangle-\langle \log{x(t)}\rangle^{2}=\sigma^{2}t.
\end{align}

\subsection{Black-Scholes formula}

As previously said, perhaps the best-known application of GBM is in finance, and in particular the BS model for pricing of European options. Formally, a European option is a contract which gives the buyer (the owner or holder of the option) the right, but not the obligation, to buy or sell an underlying asset or instrument $x(T)$ at a specified strike price $K$ on a specified date $T$. The seller has the corresponding obligation to fulfill the transaction – to sell or buy – if the buyer (owner) ``exercises'' the option. An option that conveys to the owner the right to buy at a specific price is referred to as a \textit{call}; an option that conveys the right of the owner to sell at a specific price is referred to as a \textit{put}. Here we are going to consider the valuation of call options, denoted as $C_{\text{BS}}(x,t)$, with the note that the derived results easily extend to put options.

In the modeling of financial assets, a standard assumption is that there is a risk-neutral distribution $f(x,t)$ for the price of the asset. This measure is simply a probability distribution which takes into account the risk of future price fluctuations. Once a risk-neutral distribution is assigned, the value of the option is obtained by discounting the expectation of its value at the maturity $T$ with respect to that distribution \cite{black1973pricing,merton1973theory}, i.e.,
\begin{align}
C_{\text{BS}}(x,t) = e^{-r(T - t)} \int_K^{\infty} (x(T) - K) f(x,T, |x_0,0) dx,
\label{eq:call-option}
\end{align}
where $r$ is the risk-free rate of return and $x_0$ is the asset price at the beginning ($t = 0$). Notice that the integral is calculated only for the region of prices where the option has positive value, since for asset price less than $K$ the option would not be exercised (i.e. its value is $0$).

Eqs.~\eqref{eq:call-option} and~\eqref{lognormal distribution} can be combined to derive an analytical formula for the value of the call option in the BS model for the GBM as
\begin{align}
C_{\text{BS}}(x_0,T,K,t)  &= N(d_1) x(t) - N(d_2) K e^{-(\mu-\frac{\sigma^2}{2})(T-t)} \\
d_1 &= \frac{1}{\sigma \sqrt{T-t}} \bigg[\log{\frac{x(t)}{K} + \mu(T-t)} \bigg] \\
d_2 &= d_1 - \sigma \sqrt{T-t},
    \label{eq:bs-solution}
\end{align}
where $N(x)=\frac{1}{\sqrt{2\pi}}\int_{-\infty}^{x}e^{-u^{2}/2}\,du$ is the cumulative distribution function of the Gaussian distribution with zero mean and unit variance. Put simply, the two terms in the BS formula describe the current price of the asset weighted by the probability that the investor will exercise its option at time $t$ and the discounted price of the strike price weighted by its exercise probability.  The terms $d_{1,2}$ can be seen as measures of the moneyness of the option and $N(d_{1,2})$ as probabilities that the option will expire while its value is in the money. The neat BS formulation has allowed the model to be widely applied in both theoretical investigations and empirical implementations. However, the BS model has failed to adequately reproduce a plethora of real world properties.

The European option $C_{\text{BS}}(x,t)$ (\ref{eq:call-option}) is a solution of the Black-Scholes equation, see for example \cite{hull2003options},
\begin{align}\label{bs eq}
    \left(\frac{\partial}{\partial t}+\frac{\sigma^{2}x^{2}}{2}\frac{\partial^2}{\partial x^2}-r+r\,x\frac{\partial}{\partial x}\right)C_{\text{BS}}(x,t)=0,
\end{align}
with initial condition $C_{\text{BS}}(x,T)=\max\{x-K,0\}$, $x\ge0$, and boundary conditions $C_{\text{BS}}(x=0,t)=0$, $t\geq T$, and $C_{\text{BS}}(x\rightarrow\infty,t)\rightarrow x$. By using $t=0$ and $T\rightarrow t$, one finds the equation
\begin{align}\label{bs t}
    \frac{\partial}{\partial t}C_{\text{BS}}(x,t)=\left(\frac{\sigma^{2}x^{2}}{2}\frac{\partial^2}{\partial x^2}-r+r\,x\frac{\partial}{\partial x}\right)C_{\text{BS}}(x,t).
\end{align}
with initial condition $C_{\text{BS}}(x,t=0)=\max\{x-K,0\}$, $x\ge0$, and boundary conditions $C_{\text{BS}}(x=0,t)=0$, $t\geq0$, and $C(x\rightarrow\infty,t)\rightarrow x$. 

In particular, theoretically predicted option prices with fixed values for drift $\mu$ and volatility $\sigma$ via the BS model are known to significantly deviate from their respective market values in a plethora of cases. To deal with this problem, extensions of the BS model have emerged, which include combination of the GBM with jumps~\cite{merton1976option,10}, or with stochastic volatility~\cite{11,12}.

\subsection{Subdiffusive GBM}

One of the reasons why the standard GBM is not able to explain empirical data is because it fails to explain periods of constant prices which appear on markets with low number of transactions. The price in these constant periods can be described as a trapped particle in physical systems that manifest anomalous diffusion (subdiffusion) \cite{scalas2000fractional,raberto2002waiting}. To deal with this problem, the so-called {\it subdiffusive} GBM (sGBM) has been developed~\cite{13}, by using the subordination approach. The corresponding equation for the sGBM becomes the following fractional Fokker-Planck equation~\cite{13} (see also~\cite{14})
\begin{align}\label{generalizedFPE_RL}
\frac{\partial}{\partial t}f_{\alpha}(x,t)={_{\textrm{RL}}}D_{t}^{1-\alpha}&\left[-\mu\frac{\partial}{\partial x}xf_{\alpha}(x,t)+\frac{\sigma^2}{2}\frac{\partial^2}{\partial x^2}x^{2}f_{\alpha}(x,t)\right],
\end{align}
where 
\begin{align}
{_{\textrm{RL}}}D_{t}^{\nu}f(t)=\frac{1}{\Gamma(1-\nu)}\frac{d}{dt}\int_{0}^{t}(t-t')^{-\nu}f(t')\,dt'
\end{align}
is the Riemann-Liouville fractional derivative of order $0<\nu<1$ \cite{20}\footnote{The Laplace transform of the Riemann-Liouville fractional derivative of a given function reads $\mathscr{L}\left\{{_{\textrm{RL}}}D_{t}^{\nu}f(t)\right\}(s)=s^{\nu}\mathscr{L}\left\{f(t)\right\}(s)-{_{\textrm{RL}}}I_{t}^{1-\nu}f(0+)$, where ${_{\textrm{RL}}}I_{t}^{\nu}f(t)=\frac{1}{\Gamma(\nu)}\int_{0}^{t}(t-t')^{\nu-1}f(t')\,dt'$ is the Riemann-Liouville fractional integral.}. To avoid the strange initial condition, alternatively, we could use the integral version of the equation
\begin{align}\label{generalizedFPE_RLint}
f_{\alpha}(x,t)-g_{\alpha}(x,0)={_{\textrm{RL}}}D_{t}^{-\alpha}&\left[-\mu\frac{\partial}{\partial x}xf_{\alpha}(x,t)+\frac{\sigma^2}{2}\frac{\partial^2}{\partial x^2}x^{2}f_{\alpha}(x,t)\right],
\end{align}
where $\mathscr{L}\left\{{_{\textrm{RL}}}D_{t}^{-\alpha}f(t)\right\}(s)=s^{-\alpha}\mathscr{L}\{f(t)\}(s)$. In Ref.~\cite{14} the time fractional Fokker-Planck equation~(\ref{generalizedFPE_RL}) for sGBM is derived within the CTRW theory for a particle on a geometric lattice in presence of a logarithmic potential. 

Here we note that the fractional Fokker-Planck equation~(\ref{generalizedFPE_RL}) can be obtained by using the Langevin equation approach \cite{fogedby1994langevin}, i.e., by considering a CTRW model described by a coupled Langevin equations \cite{li2016option},
\begin{align}
    & \frac{d}{du}x(u)=\mu\,x(u)+\sigma\,x(u)\,\xi(u), \label{coupled le1} \\
    & \frac{d}{du}\mathcal{T}(u)=\zeta(u). \label{coupled le2}
\end{align}
Therefore, $x(t)$ is parametrised in terms of the number of steps $u$, and the connection to the physical time $t$ is given by $\mathcal{T}(u)=\int_{0}^{u}\tau(u')\,du$, where $\tau(u)$ is a total of individual waiting times $\tau$ for each step. In mathematical terms this is called subordination \cite{feller2008introduction,magdziarz2007fractional,magdziarz2008equivalence}. The noise $\xi(u)$ is a white noise with zero mean and correlation $\langle \xi(u)\xi(u')\rangle=2\delta(u-u')$, while $\zeta(u)$ is one-sided $\alpha$-stable L\'evy noise with the stable index $0<\alpha<1$. The inverse process $\mathcal{S}(t)$ of the one-sided $\alpha$-stable Levy process $\mathcal{T}(u)$ with characteristic function $\langle e^{-s\mathcal{T}(u)}\rangle=e^{-s^{\alpha}u}$ is given by $\mathcal{S}(t)=\mathrm{inf}\left\{u>0: \mathcal{T}(u)>t\right\}$, i.e., it represents a collection of first passage times \cite{fogedby1994langevin}. The CTRW is defined by the subordinated process $\mathcal{X}(t)=x(\mathcal{S}(t))$. The PDF $h(u,t) $ of the inverse process $\mathcal{S}(t)$ can be found from the relation \cite{fogedby1994langevin}
\begin{align}
    h(u,t)=-\frac{\partial}{\partial u}\Theta\left(t-\mathcal{T}(u)\right),
\end{align}
where $\Theta(z)$ is the Heaviside theta function. The Laplace transform then yields
\begin{align}
    \hat{h}(u,s)=-\frac{\partial}{\partial u}\frac{1}{s}\,\left\langle\int_{0}^{\infty}\delta\left(t-\mathcal{T}(u)\right)e^{-st}\,dt\right\rangle=-\frac{\partial}{\partial u}\frac{1}{s}\langle e^{-s\mathcal{T}(u)}\rangle=-\frac{\partial}{\partial u}\frac{1}{s}e^{-s^{\alpha}u}=s^{\alpha-1}e^{-s^{\alpha}u}.
\end{align}
Therefore, $f_{\alpha}(x,t)=\langle\delta(x-\mathcal{X}(t))\rangle=\langle\delta(x-X(\mathcal{S}(t))\rangle=\int_{0}^{\infty}f(x,u)\,h(u,t)\,dt$, from where one can easily arrive to the fractional Fokker-Planck equation~(\ref{generalizedFPE_RL}).

The mean value for sGBM is given by~\cite{14,li2016option}
\begin{align}\label{mean_power}
\langle x(t)\rangle = x_0\, E_{\alpha}\left(\mu t^{\alpha}\right),
\end{align}
where $E_{\alpha}(z)$ is the one parameter Mittag-Leffler (ML) function \cite{20,scalas2000fractional}\footnote{The Laplace transform of the one parameter ML function reads $\mathscr{L}\left\{E_{\alpha}(at^{\alpha})\right\}(s)=\frac{s^{\alpha-1}}{s^{\alpha}-a}$.}
\begin{align}\label{one parameter ml}
    E_{\alpha}(z)=\sum_{k=0}^{\infty}\frac{z^{k}}{\Gamma(\alpha k+1)},
\end{align}
with $(z \in C; \Re(\alpha) > 0)$, and $\Gamma(\cdot)$ is the Gamma function. The ML function is a generalization of the exponential function since $E_{1}(z)=e^{z}$. The asymptotic behavior of the mean is given by\footnote{For the short time limit we use the first two terms from the series expansion of the ML function~(\ref{one parameter ml}), while for the long time limit we apply its asymptotic expansion formula \cite{20,garra2018prabhakar}, $E_{\alpha}(z)\simeq\frac{1}{\alpha}e^{z^{1/\alpha}}$, $z\gg1$. Here we note that the asymptotic behavior of the ML function with negative argument has a power-law form, i.e., $E_{\alpha}(-z^{\alpha})\simeq\frac{z^{-\alpha}}{\Gamma(1-\alpha)}$ for $z\ll1$ and $0<\alpha<2$ \cite{20,garra2018prabhakar}.}
\begin{align}
\left\langle x(t)\right\rangle\simeq
x_0\,
\left\lbrace\begin{array}{l l l}
     \smallskip & 1+\mu t^{\alpha}/\Gamma(1+\alpha)\sim e^{\mu t^{\alpha}/\Gamma(1+\alpha)}, \quad & t\ll1, \\
     & {\alpha^{-1}}e^{\mu^{1/\alpha}t}, \quad & t\gg1.
\end{array}\right.
\end{align}
The MSD also is given through the one parameter ML function \cite{14,li2016option}
\begin{align}
\langle x^{2}(t)\rangle 
&=x_{0}^{2}\, E_{\alpha}\left((\sigma^{2}+2\mu) t^{\alpha}\right)\nonumber\\&\simeq 
\langle x^2(0)\rangle
\left\lbrace\begin{array}{l l l}
     \smallskip & 1+(\sigma^{2}+2\mu) t^{\alpha}/\Gamma(1+\alpha)\sim e^{(\sigma^{2}+2\mu) t^{\alpha}/\Gamma(1+\alpha)}, \quad & t\ll1, \\
     & {\alpha^{-1}}e^{(\sigma^2+2\mu)^{1/\alpha}t}, \quad & t\gg1.
\end{array}\right.
\end{align}
From here one concludes that the sGBM is an exponentially fast process.

The first log-moment has the form \cite{14}
\begin{align}
    \langle \log{x(t)}\rangle=\langle \log{x_0}\rangle+\bar{\mu}\,\int_{0}^{t}\frac{t'^{\alpha-1}}{\Gamma(\alpha)}\,dt'=\langle \log{x_0}\rangle+\bar{\mu}\,\frac{t^{\alpha}}{\Gamma(1+\alpha)},
\end{align}
which gives a power-law dependence with respect to time of the expectation of the log return with period $\Delta{t}$, i.e., \cite{14}
\begin{align}
    \frac{1}{\Delta{t}}\langle \log{\left(x(t+\Delta{t})/x(t)\right)}\rangle \underset{\Delta{t}\rightarrow0}{\sim}\bar{\mu}\,\frac{t^{\alpha-1}}{\Gamma(\alpha)}.
\end{align}
Such models have been used, for example, to explain the dynamics of an asset before a market crash~\cite{sornette1996stock}. The second log-moment becomes \cite{14}
\begin{align}
    \langle \log^{2}{x(t)}\rangle 
    =\langle \log^{2}{x_0}\rangle+\left[2\bar{\mu}\langle\log{x_0}\rangle+\sigma^{2}\right]\frac{t^{\alpha}}{\Gamma(1+\alpha)}+2\bar{\mu}^{2}\frac{t^{2\alpha}}{\Gamma(1+2\alpha)}.
\end{align}
from where for the log-variance one finds \cite{14}
\begin{align}
    \langle \log^{2}{x(t)}\rangle-\langle \log{x(t)}\rangle^2=\sigma^{2}\frac{t^{\alpha}}{\Gamma(1+\alpha)}+\bar{\mu}^{2}\left[\frac{2}{\Gamma(1+2\alpha)}-\frac{1}{\Gamma^{2}(1+\alpha)}\right]t^{2\alpha},
\end{align}
which in the long time limit scales as $t^{2\alpha}$ ($0<\alpha<1$), contrary to the linear scaling $t$ for regular GBM ($\alpha=1$).


\section{Generalised GBM}\label{sec:gen-gbm}

In this section we consider a generalization of GBM, under which the standard and subdiffusive GBM arise as special cases, by using the subordination approach. The continuous time random walk approach to the corresponding Fokker-Planck equation is given in Appendix~\ref{sec:gen-gbm.2} in detail.

The same Fokker-Planck equation can be obtained by using the coupled Langevin equations approach \cite{fogedby1994langevin}, as given by Eqs. (\ref{coupled le1}) and (\ref{coupled le2}), where the waiting times are given by $\langle e^{-s\mathcal{T}(u)}\rangle=e^{-\hat{\Psi}(s)u}$, with $\hat{\Psi}(s)=1/\hat{\eta}(s)$.

\subsection{
Subordination approach}\label{sec:gen-gbm.1}

The generalisation of GBM which we consider is in the form of the stochastic process 
\begin{align}
\mathcal{X}(t)=x\left(\mathcal{S}(t)\right),
\end{align}
where $\mathcal{X}(t)$ is the {\it generalised} GBM (gGBM)\footnote{The current process should not be confused with the Pagnini-Mainardi generalised grey Brownian motion, see Ref.~\cite{mura2008non,mura2008characterizations,sposini2018random}.}, $\mathcal{S}(t)=\mathrm{inf}\left\{u>0: \mathcal{T}(u)>t\right\}$ is the operational time, and $\mathcal{T}(u)$ is an infinite divisible process, i.e., a strictly increasing L\'evy motion with $$\langle e^{-s\mathcal{T}(u)}\rangle=e^{-u\hat{\Psi}(s)},$$ and $\hat{\Psi}(s)$ is the L\'evy exponent \cite{5,13,15}. Here we consider $\hat{\Psi}(s)=1/\hat{\eta}(s)$.

Next we find the PDF of gGBM which subordinates the processes from the time scale $t$ (physical time) to the GBM on a time scale $u$ (operational time). Therefore, the PDF $P(x,t)$ of a given random process $\mathcal{X}(t)$ can be represented as \cite{13,16,17,18,schulz2014aging}
\begin{align}\label{sub}
P(x,t)=\int_{0}^{\infty}f(x,u)h(u,t)\,du,
\end{align}
where $f(x,u)$ satisfies the Fokker-Planck equation~(\ref{generalizedFPE_standard}) for the standard GBM. The function $h(u, t)$ is the PDF subordinating the random process $\mathcal{X}(t)$ to the standard GBM. In the Laplace space, Eq.~(\ref{sub}) reads
\begin{align}\label{sub laplace}
\hat{P}(x,s)=\mathscr{L}\left\{P(x,t)\right\}&=\int_{0}^{\infty}e^{-st}P(x,t)\,dt=\int_{0}^{\infty}f(x,u)\hat{h}(u,s)\,du,
\end{align}
where $\hat{h}(u,s)=\mathscr{L}\left\{h(u,t)\right\}$. By considering 
\begin{align}\label{subordination function}
\hat{h}(u,s)=\frac{\hat{\Psi}(s)}{s}e^{-u\hat{\Psi}(s)}=\frac{1}{s\hat{\eta}(s)}e^{-\frac{u}{\hat{\eta}(s)}},
\end{align}
we then have
\begin{align}\label{sub laplace 2}
\hat{P}(x,s)=\frac{1}{s\hat{\eta}(s)}\int_{0}^{\infty}f(x,u)e^{-\frac{u}{\hat{\eta}(s)}}\,du=\frac{1}{s\hat{\eta}(s)}\,\hat{f}\left(x,\frac{1}{\hat{\eta}(s)}\right).
\end{align}
By Laplace transform of the Fokker-Planck equation (\ref{generalizedFPE_standard}) for the GBM, and using relation (\ref{sub laplace 2}), one finds that the PDF $P(x,s)$ satisfies
\begin{align}\label{generalizedFPE2L}
s\hat{P}(x,s)-P(x,0)=s\,\hat{\eta}(s)&\left[-\mu\frac{\partial}{\partial x}x\hat{P}(x,s)+\frac{\sigma^2}{2}\frac{\partial^2}{\partial x^2}x^{2}\hat{P}(x,s)\right].
\end{align}
After inverse Laplace transform we arrive at the generalised Fokker-Planck equation (see Refs.~\cite{15,li2016option} where one-sided $\alpha$-stable waiting times are considered in detail) 
\begin{align}\label{generalizedFPE2}
\frac{\partial}{\partial t}P(x,t)=\frac{\partial}{\partial t}\int_{0}^{t}\eta(t-t')&\left[-\mu\frac{\partial}{\partial x}xP(x,t')+\frac{\sigma^2}{2}\frac{\partial^2}{\partial x^2}x^{2}P(x,t')\right]dt',
\end{align}
where $\eta(t)$ is a so-called memory kernel. One observes that for $\eta(t)=1$ we arrive at the Fokker-Planck equation (\ref{generalizedFPE_standard}) for the GBM, and for $\eta(t)=\frac{t^{\alpha-1}}{\Gamma(\alpha)}$ at the time fractional Fokker-Planck equation (\ref{generalizedFPE_RL}) for the sGBM. From Eqs.~(\ref{sub laplace}) and (\ref{subordination function}), we find for the PDF in the Laplace domain, see also Ref.~\cite{li2016option},
\begin{align}\label{sub general}
\hat{P}(x,s)&=\int_{0}^{\infty}\frac{1}{x \sqrt{2\pi\sigma^{2}u}}\times\exp\bigg( - \frac{\big[\log{x} - \log{x_0} - \bar{\mu} u\big]^2}{2\sigma^2 u}\bigg)\frac{1}{s\hat{\eta}(s)}e^{-\frac{u}{\hat{\eta}(s)}}\,du\nonumber\\&=\frac{1/[s\hat{\eta}(s)]}{x\sqrt{\bar{\mu}^{2}+2\sigma^{2}/\hat{\eta}(s)}}\left\lbrace\begin{array}{l l l}
     \smallskip & \exp\left(-\frac{\log{x}-\log{x_0}}{\sigma^{2}}\left[\sqrt{\bar{\mu}^{2}+2\sigma^{2}/\hat{\eta}(s)}-\bar{\mu}\right]\right), \quad & x>x_0, \\ \smallskip
     & 1, \quad & x=x_0, \\
     & \exp\left(\frac{\log{x}-\log{x_0}}{\sigma^{2}}\left[\sqrt{\bar{\mu}^{2}+2\sigma^{2}/\hat{\eta}(s)}+\bar{\mu}\right]\right), \quad & x<x_0,
\end{array}\right.
\end{align}

\paragraph{{\bf Remark 1:}} Here we note that there are restrictions on the choice of the memory kernel $\eta(t)$ since the PDF~(\ref{sub}) should be non-negative. From the subordination integral it follows that the subordination function $h(u,t)$ should be non-negative, which, according to the Bernstein theorem, means that its Laplace transform~(\ref{subordination function}) should be a completely monotone function \cite{schilling2012bernstein}. Therefore, the PDF~(\ref{sub}) will be non-negative if $1/[s\hat{\eta}(s)]$ is a completely monotone function, and $1/\hat{\eta}(s)$ is a Bernstein function, see Refs.~\cite{19,sandev2017beyond}.

\paragraph{{\bf Remark 2:}}We note that Eq.~(\ref{generalizedFPE2L}) can be written in an equivalent form as
\begin{align}\label{generalizedFPE}
\int_{0}^{t}\gamma(t-t')\frac{\partial}{\partial t'}P(x,t')\,dt'&=-\mu\frac{\partial}{\partial x}xP(x,t)+\frac{\sigma^2}{2}\frac{\partial^2}{\partial x^2}x^{2}P(x,t),
\end{align}
where the memory kernel $\gamma(t)$ is connected to $\eta(t)$ in Laplace space as $\gamma(s)=1/[s\eta(s)]$ \cite{19}. From this relation we find that for GBM ($\eta(t)=1$, i.e., $\hat{\eta}(s)=1/s$) the memory kernel $\gamma(t)$ is given by $\gamma(t)=\mathscr{L}^{-1}\left\{s^{-1}\hat{\eta}^{-1}(s)\right\}=\mathscr{L}^{-1}\left\{1\right\}=\delta(t)$. For sGBM ($\eta(t)=t^{\alpha-1}/\Gamma(\alpha)$, i.e., $\hat{\eta}(s)=s^{-\alpha}$) the memory kernel becomes $\gamma(t)=\mathscr{L}^{-1}\left\{s^{\alpha-1}\right\}=t^{-\alpha}/\Gamma(1-\alpha)$, and thus Eq.~(\ref{generalizedFPE}) reads
\begin{align}\label{caputoFPE2}
{_{\textrm{C}}}D_{t}^{\alpha}P(x,t)&=-\mu\frac{\partial}{\partial x}xP(x,t)+\frac{\sigma^2}{2}\frac{\partial^2}{\partial x^2}x^{2}P(x,t),
\end{align}
where
\begin{align}
{_{\textrm{C}}}D_{t}^{\nu}f(t)=\frac{1}{\Gamma(1-\nu)}\int_{0}^{t}(t-t')^{-\nu}\frac{d}{dt'}f(t')\,dt'
\end{align}
is the Caputo fractional derivative of order $0<\nu<1$ \cite{20}\footnote{The Laplace transform of the Caputo derivative of a given function reads $\mathscr{L}\left\{{_{\textrm{C}}}D_{t}^{\nu}f(t)\right\}(s)=s^{\nu}\mathscr{L}\left\{f(t)\right\}(s)-s^{\nu-1}f(0+)$.}. We note that with the appropriate restrictions for $\eta(t)$ and $\gamma(t)$ both formulations are equivalent.  

\paragraph{{\bf Remark 3:}}
For $\eta(t)=\frac{t^{\alpha-1}}{\Gamma(\alpha)}$, $0<\alpha<1$, gGBM corresponds to sGBM. From the subordination approach one finds \cite{13}
\begin{eqnarray}
\hat{h}(u,s)=s^{\alpha-1}e^{-us^{\alpha}}=s^{\alpha-1} H_{0,1}^{1,0}\left[u\,s^{\alpha}\left|\begin{array}{l l}
    - \\
    (0,1)
  \end{array}\right.\right], 
\end{eqnarray}
where $H_{p,q}^{m,n}(z)$ is the Fox $H$-function, see Appendix~\ref{appH}. By inverse Laplace transform (\ref{h laplace}) we obtained \cite{13}
\begin{align}
h(u,t)=\mathscr{L}^{-1}\left\{\hat{h}(u,s)\right\}
=t^{-\alpha}\,H_{1,1}^{1,0}\left[\frac{u}{t^{\alpha}}\left|\begin{array}{l l}
    (1-\alpha,\alpha) \\
    (0,1)
  \end{array}\right.\right]
=\frac{1}{u}\,H_{1,1}^{1,0}\left[\frac{u}{t^{\alpha}}\left|\begin{array}{l l}
    (1,\alpha) \\
    (1,1)
  \end{array}\right.\right],
\end{align}
where we applied property~(\ref{h pr1}). The solution in Laplace space then becomes
\begin{align}\label{sub2}
\hat{P}(x,s)&=\int_{0}^{\infty}\frac{1}{x \sqrt{2\pi\sigma^{2}u}}\times\exp\bigg( - \frac{\big[\log{x} - \log{x_0} - \bar{\mu} u\big]^2}{2\sigma^2 u}\bigg)s^{\alpha-1}e^{-us^{\alpha}}\,du\nonumber\\&=\frac{s^{\alpha-1}}{x\sqrt{\bar{\mu}^{2}+2\sigma^{2}s^{\alpha}}}\times\left\lbrace\begin{array}{l l l}
     \smallskip & \exp\left(-\frac{\log{x}-\log{x_0}}{\sigma^{2}}\left[\sqrt{\bar{\mu}^{2}+2\sigma^{2}s^{\alpha}}-\bar{\mu}\right]\right), \quad & x>x_0, \\
     \smallskip & 1, \quad & x=x_0, \\
     & \exp\left(\frac{\log{x}-\log{x_0}}{\sigma^{2}}\left[\sqrt{\bar{\mu}^{2}+2\sigma^{2}s^{\alpha}}+\bar{\mu}\right]\right), \quad & x<x_0,
\end{array}\right.
\end{align}
which is obtained in Ref.~\cite{li2016option} in a similar way. From here we can plot the PDF by using numerical inverse Laplace transform techniques.

\subsection{Generalised BS formula}

If we consider that the asset price follows a gGBM, the generalised BS (gBS) formula for the option price is \cite{15}
\begin{align}
    C_{\text{gBS}}(x,t)=\langle e^{-r(\mathcal{S}(T)-t)}(x(\mathcal{S}(T))-K)\rangle_{x}=\int_{0}^{\infty}C_{\text{BS}}(x,u)\,h(u,T)\,du,
\end{align}
where $C_{\text{BS}}(x,t)$ is taken from the BS formula (\ref{eq:bs-solution}), and $h(x,T)$ is the subordination function defined by Eq.~(\ref{subordination function}) in the Laplace domain. By Laplace transform one finds
\begin{align}
    \hat{C}_{\text{gBS}}(x,s)=\frac{1}{s\hat{\eta}(s)}\,\hat{C}_{\text{BS}}(x,1/\hat{\eta}(s)).
\end{align}
Therefore, from Eq.~(\ref{bs t}), the corresponding equation for the option price becomes \cite{li2016option}
\begin{align}
    \frac{\partial}{\partial t}C_{\text{gBS}}(x,t)=\frac{\partial}{\partial t}\int_{0}^{t}\eta(t-t')\left(\frac{\sigma^{2}x^{2}}{2}\frac{\partial^2}{\partial x^2}-r+r\,x\frac{\partial}{\partial x}\right)C_{\text{gBS}}(x,t')\,dt.
\end{align}

\subsection{Calculation of moments}

The $n$th moment $\langle \mathcal{X}^{n}(t)\rangle=\int_{0}^{\infty}x^{n}\,P(x,t)\,dx$ can be calculated by multiplying both sides of Eq.~(\ref{generalizedFPE2}) by $x^n$ and integration over $x$, see Appendix~\ref{app3}. In the Laplace domain, this results in
\begin{align}\label{n-th moment laplace}
\langle \hat{\mathcal{X}}^{n}(s)\rangle=\frac{s^{-1}}{1-\hat{\eta}(s)\left[\frac{\sigma^{2}}{2}n(n-1)+\mu\,n\right]}\langle x_{0}^{n}\rangle.
\end{align}
From this result we reproduce the normalization condition $\langle x^{0}(t)\rangle=\langle x_{0}^{0}\rangle=1$. The general results for the mean value ($n=1$) and the MSD ($n=2$) in terms of the memory kernel become \cite{li2016option},
\begin{align}\label{mean general eta}
\langle \hat{\mathcal{X}}(s)\rangle=x_0\,\frac{s^{-1}}{1-\mu\hat{\eta}(s)},
\end{align}
and
\begin{align}
\langle \hat{\mathcal{X}}^2(s)\rangle=x_{0}^{2}\,\frac{s^{-1}}{1-(\sigma^{2}+2\mu)\hat{\eta}(s)}.
\end{align}

The log-moments $\left\langle \log^{n}{x(t)}\right\rangle=\int_{0}^{\infty}\log^{n}x\,P(x,t)\,dx$, can also be calculated exactly through the memory kernel, see Appendix~\ref{app3}. The normalization condition is satisfied, i.e., $\langle \log^{0}{x(t)}\rangle=1$, while the log-mean reads
\begin{align}
    \langle \log{x(t)}\rangle=\langle \log{x_0}\rangle+\bar{\mu}\int_{0}^{t}\eta(t')\,dt'.
\end{align}
From here, we find for the expectation of the periodic log return with period $\Delta{t}$
\begin{align}
    \frac{1}{\Delta{t}}\langle \log{\left(x(t+\Delta{t})/x(t)\right)}\rangle&=\bar{\mu}\frac{1}{\Delta{t}}\int_{t}^{t+\Delta{t}}\eta(t')\,dt'\nonumber\\&=\bar{\mu}\frac{I(t+\Delta t)-I(t)}{\Delta t}\underset{\Delta{t}\rightarrow0}{\sim}\bar{\mu}\,\eta(t),
\end{align}
where $I(t)=\int\eta(t)\,dt$, i.e., $I'(t)=\eta(t)$. Therefore, the expectation of the periodic log returns behaves as the rate of the first log-moment,
\begin{align}
    \frac{1}{\Delta{t}}\langle \log{\left(x(t+\Delta{t})/x(t)\right)}\rangle\underset{\Delta{t}\rightarrow0}{\sim}\frac{d}{dt}\langle\log{x(t)}\rangle,
\end{align}
which is proportional to the memory kernel $\eta(t)$. Moreover, for the second log-moment we find
\begin{align}
    \langle \log^{2}{x(t)}\rangle&=\langle \log^{2}{x_0}\rangle\nonumber\\&+\int_{0}^{t}\eta(t-t')\left\{2\bar{\mu}\left[\langle \log{x_0}\rangle+\bar{\mu}\int_{0}^{t'}\eta(t'')\,dt''\right]+\sigma^{2}\right\}dt',
\end{align}
from where the log-variance becomes
\begin{align}
    &\langle \log^{2}{x(t)}\rangle-\langle \log{x(t)}\rangle^{2}\nonumber\\&=\sigma^{2}\int_{0}^{t}\eta(t')\,dt'+\bar{\mu}^{2}\left[2\int_{0}^{t}\eta(t-t')\left(\int_{0}^{t'}\eta(t'')\,dt''\right)dt'-\left(\int_{0}^{t}\eta(t')\,dt'\right)^2\right].
\end{align}

From all these general formulas one can easily recover the previous results for the standard GBM ($\eta(t)=1$, i.e., $\hat{\eta}(s)=1/s$) and sGBM ($\eta(t)=t^{\alpha-1}/\Gamma(\alpha)$, i.e., $\hat{\eta}(s)=s^{-\alpha}$, $0<\alpha<1$).

\subsection{Exponentially truncated subdiffusive GBM}\label{sec:gen-gbm.5}

As an example for another memory kernel in gGBM we consider a power-law memory kernel with exponential truncation,
\begin{align}
\eta(t)=\frac{t^{\alpha-1}}{\Gamma(\alpha)}e^{-\frac{t}{\tau}},\end{align}
where $\tau$ is a characteristic crossover time scale, $ 0 < \alpha < 1$. Such forms are important in many real-world applications, in which the scale-free nature of the waiting time dynamics is broken at macroscopic times $t\gg\tau$ \cite{19}. Therefore,
\begin{align}
\hat{\eta}(s)=(s+\tau^{-1})^{-\alpha},
\end{align}
where we use the shift rule of the Laplace transform, $\mathscr{L}\left\{e^{-at}f(t)\right\}=\hat{F}(s+a)$, for $\hat{F}(s)=\mathscr{L}\left\{f(t)\right\}$.

The mean value reads,
\begin{align}
\left\langle x(t)\right\rangle&=x_0\,\mathscr{L}^{-1}\left\{\frac{s^{-1}}{1-\mu(s+\tau^{-1})^{-\alpha}}\right\}(t)=x_0\,\int_{0}^{t}e^{-t'/\tau}t'^{-1}E_{\alpha,0}\left(\mu t'^{\alpha}\right),
\end{align}
and the MSD
\begin{align}
\langle x^2(t)\rangle=x_{0}^{2}\,\int_{0}^{t}e^{-t'/\tau}t'^{-1}E_{\alpha,0}\left((\sigma^{2}+2\mu) t'^{\alpha}\right),
\end{align}
where
\begin{align}
    E_{\alpha,\beta}(z)=\sum_{k=0}^{\infty}\frac{z^{k}}{\Gamma(\alpha k+\beta)}
\end{align}
$(z,\beta \in C; \Re(\alpha) > 0)$ is the two parameter ML function \cite{20}\footnote{The Laplace transform of the two parameter ML function reads
$\mathscr{L}\left\{t^{\beta-1}E_{\alpha,\beta}(at^{\alpha})\right\}(s)=\frac{s^{\alpha-\beta}}{s^{\alpha}-a}$.}. From here, for the short time limit we obtain the results for the sGBM
\begin{align}
\langle x(t)\rangle\simeq x_0\, E_{\alpha}\left(\mu t^{\alpha}\right),
\end{align}
\begin{align}
\langle x^2(t)\rangle\simeq x_{0}^{2}\, E_{\alpha}\left((\sigma^{2}+2\mu) t^{\alpha}\right),
\end{align}
since the exponential truncation has no effect for short times, $e^{-t/\tau}\simeq1-t/\tau$, $t/\tau\ll1$. The long time limit ($s\tau\ll1$) yields\footnote{Here we use the asymptotic expansion formula for the two parameter ML function
$E_{\alpha,\beta}(z)\simeq\frac{1}{\alpha}\,e^{z^{1/\alpha}}z^{(1-\beta)/\alpha}$, $z\gg1$ \cite{20,garra2018prabhakar}. Here we note that the asymptotic behavior for negative arguments is given by power-law decay, $E_{\alpha,\beta}\left(-z^{\alpha}\right)\simeq\frac{z^{-\alpha}}{\Gamma(\beta-\alpha)}$, $z\gg1$ \cite{20,garra2018prabhakar}.}
\begin{align}
\langle x(t)\rangle\simeq x_0\,\frac{\mu^{1/\alpha}}{\alpha(\mu^{1/\alpha}-\tau^{-1})}\left[e^{(\mu^{1/\alpha}-\tau^{-1})t}-1\right],
\end{align}
and
\begin{align}
\langle x^2(t)\rangle\simeq x_{0}^{2}\,\frac{(\sigma^{2}+2\mu)^{\frac{1}{\alpha}}}{\alpha\left([\sigma^{2}+2\mu]^{1/\alpha}-\tau^{-1}\right)}\left\{e^{\left([\sigma^{2}-2\mu]^{1/\alpha}-\tau^{-1}\right)t}-1\right\}.
\end{align}

From the general result for the log-mean, we find
\begin{align}
    \langle \log{x(t)}\rangle&=\langle \log{x_0}\rangle+\bar{\mu}\,e^{-t/\tau}\,t^{\alpha}E_{1,\alpha+1}(t/\tau)=\langle \log{x_0}\rangle+\bar{\mu}\,\tau^{\alpha}\frac{\gamma(\alpha,t/\tau)}{\Gamma(\alpha)},
\end{align}
where $\gamma(a,z)=\int_{0}^{z}t^{a-1}e^{-t}\,dt=\Gamma(a)e^{-z}\,z^{a}E_{1,a+1}(z)$ is the incomplete gamma function. For the expectation of the periodic log return with period $\Delta{t}$ we find
\begin{align}
    \frac{1}{\Delta{t}}\langle \log{\left(x(t+\Delta{t})/x(t)\right)}\rangle\underset{\Delta{t}\rightarrow0}{\sim}\bar{\mu}\,\frac{t^{\alpha-1}}{\Gamma(\alpha)}e^{-t/\tau}=\frac{d}{dt}\langle\log{x(t)}\rangle.
\end{align}
This leads to a long run log return of 0, whereas on the short time scale the same observable behaves in the same way as sGBM. As such, the model can be used to model early herd behavior where the price of an asset grows simply as a consequence of investors following trends (short run behavior), that last until the trade of the asset becomes congested (long run behavior). The second log-moment is
\begin{align}
    \langle \log^{2}{x(t)}\rangle&=\langle \log^{2}{x_0}\rangle+\left[2\bar{\mu}\langle\log{x_0}\rangle+\sigma^{2}\right]e^{-t/\tau}t^{\alpha}E_{1,\alpha+1}(t/\tau)+2\bar{\mu}^{2}e^{-t/\tau}\,t^{2\alpha}E_{1,2\alpha+1}(t/\tau),
\end{align}
from where the log-variance becomes
\begin{align}
    \langle \log^{2}{x(t)}\rangle-\langle \log{x(t)}\rangle^{2}&=2\bar{\mu}\,e^{-t/\tau}\,t^{2\alpha}E_{1,2\alpha+1}(t/\tau)\nonumber\\&
    +e^{-t/\tau}\,t^{\alpha}E_{1,\alpha+1}(t/\tau)\left[\sigma^{2}-\bar{\mu}\,e^{-t/\tau}\,t^{\alpha}E_{1,\alpha+1}(t/\tau)\right].
\end{align}
Here we note that for $t/\tau\ll1$, the obtained results correspond to those obtained for sGBM, as it should be since the exponential truncation has no influence on the process. We observe that on the long run the log variance becomes constant, i.e., it is equal to $\sigma^{2}\tau^{\alpha}+\bar{\mu}\,\tau^{2\alpha}$.

The subordination function in this case is given by
\begin{align}
&\hat{h}(u,s)=\frac{(s+\tau^{-1})^{\alpha}}{s}e^{-u(s+\tau^{-1})^{\alpha}}=\left[1+(s\tau)^{-1}\right](s+\tau^{-1})^{\alpha-1}e^{-u(s+\tau^{-1})^{\alpha}},\nonumber\\
&h(u,t)=e^{-t/\tau}\,H_{1,1}^{1,0}\left[\frac{u}{t^{\alpha}}\left|\begin{array}{c l}
    (1,\alpha) \\
    (1,1)
  \end{array}\right.\right]+\frac{1}{\tau}\int_{0}^{t}e^{-t'/\tau}\,H_{1,1}^{1,0}\left[\frac{u}{t'^{\alpha}}\left|\begin{array}{c l}
    (1,\alpha) \\
    (1,1)
  \end{array}\right.\right]dt',
\end{align}
from where one can analyze the PDF $P(x,t)$.

\subsection{Combined standard and subdiffusive GBM}\label{sec:gen-gbm.6}

As another application, let us consider the combination of GBM and sGBM, represented by the memory kernel
\begin{align}
\eta(t)=w_{1}\frac{t^{\alpha-1}}{\Gamma(\alpha)}+w_{2},
\end{align}
where $0<\alpha<1$, $w_{1}+w_{2}=1$, and 
\begin{align}
\hat{\eta}(s)=w_{1}s^{-\alpha}+w_{2}s^{-1}.
\end{align}
This case combines both motions governed by Eq.~(\ref{generalizedFPE_standard}) and (\ref{generalizedFPE_RL}). In this case, in a jump picture normal GBM steps occur with weight $w_{2}$ while power-law waiting time steps are realised with weight $w_{1}$. 

The mean value for this case is given by
\begin{align}\label{GBM and FGBM mean}
\left\langle x(t)\right\rangle&=x_0\,\mathscr{L}^{-1}\left\{\frac{s^{-1}}{1-\mu\left(w_{1}s^{-\alpha}+w_{2}s^{-1}\right)}\right\}(t)=x_0\,\sum_{n=0}^{\infty}w_{1}^{n}\mu^{n}t^{\alpha n}E_{1,\alpha n+1}^{n+1}\left(w_{2}\mu t\right)\nonumber\\&= x_0\,\sum_{n=0}^{\infty}\frac{w_{1}^{n}\mu^{n}t^{\alpha n}}{\Gamma(\alpha n+1)}{_1}F_{1}\left(n+1;\alpha n+1;w_{2}\mu t\right),
\end{align}
where ${_1}F_{1}\left(a;b;z\right)=\sum_{k=0}^{\infty}\frac{(a)_{k}}{(b)_{k}}\frac{z^{k}}{k!}$ is the Kummer confluent hypergeometric function, and
\begin{align}
E_{\alpha,\beta}^{\gamma}(z)=\sum_{n=0}^{\infty}\frac{(\gamma)_{n}}{\Gamma(\alpha n+\beta)}\frac{z^{n}}{n!},
\end{align}
is the three parameter ML function \cite{prabhakar}, $(\gamma)_{n}=\Gamma(\gamma+n)/\Gamma(\gamma)$ is the Pochhammer symbol\footnote{The Laplace transform of the three parameter ML function reads
$\mathscr{L}\left\{t^{\beta-1}E_{\alpha,\beta}^{\gamma}(at^{\alpha})\right\}(s)=\frac{s^{\alpha\gamma-\beta}}{\left(s^{\alpha}-a\right)^{\gamma}}$.}. From here we see that for $w_1=0$ and $w_2=1$ only the term for $n=0$ in Eq.~(\ref{GBM and FGBM mean}) survives which yields the result for standard GBM as it should be. The opposite case, with $w_1=1$ and $w_2=0$, yields 
\begin{align}
\left\langle x(t)\right\rangle&=x_0\,\sum_{n=0}^{\infty}\frac{\mu^{n}t^{\alpha n}}{\Gamma(\alpha n+1)}=E_{\alpha}\left(\mu t^{\alpha}\right),
\end{align}
as it should be for the sGBM. For the second moment we find
\begin{align}
\langle x^2(t)\rangle=x_{0}^{2}\,\sum_{n=0}^{\infty}w_{1}^{n}(\sigma^{2}+2\mu)^{n}t^{\alpha n}E_{1,\alpha n+1}^{n+1}\left(w_{2}(\sigma^{2}+2\mu)t\right).
\end{align}

Following the same procedure as previous, for the log-mean we find
\begin{align}
    \langle \log{x(t)}\rangle=\langle \log{x_0}\rangle+\bar{\mu}\left[w_1\,\frac{t^{\alpha}}{\Gamma(\alpha+1)}+w_2\,t\right].
\end{align}
and for the expectation of the periodic log return with period $\Delta{t}$, 
\begin{align}
    \frac{1}{\Delta{t}}\langle \log{\left(x(t+\Delta{t})/x(t)\right)}\rangle\underset{\Delta{t}\rightarrow0}{\sim}\bar{\mu}\left[w_1\,\frac{t^{\alpha-1}}{\Gamma(\alpha)}+w_2\right]=\frac{d}{dt}\langle\log{x(t)}\rangle.
\end{align}
This model introduces subdiffusive and trapping asset dynamics on short time scales (i.e., then the part multiplied with $w_1$ is much bigger), whereas on the long run we recover the standard GBM dynamics.
The second log-moment yields
\begin{align}
    \langle \log^{2}{x(t)}\rangle&=\langle \log^{2}{x_0}\rangle+2\bar{\mu}^{2}\left[\frac{w_{1}^{2}t^{2\alpha}}{\Gamma(2\alpha+1)}+\frac{2w_{1}w_{2}t^{\alpha+1}}{\Gamma(\alpha+2)}+\frac{w_{2}^{2}t^{2}}{2}\right]\nonumber\\&+\left\{2\bar{\mu}\langle \log{x_0}\rangle+\sigma^{2}\right\}\left(w_{1}\,\frac{t^{\alpha}}{\Gamma(\alpha+1)}+w_{2}\,t\right),
\end{align}
from where the log-variance becomes
\begin{align}
    \langle \log^{2}{x(t)}\rangle-\langle \log{x(t)}\rangle^{2}&=\sigma^{2}\left(w_{1}\,\frac{t^{\alpha}}{\Gamma(\alpha+1)}+w_{2}\,t\right)+\bar{\mu}^{2}w_{1}^{2}t^{2\alpha}\left(\frac{2}{\Gamma(2\alpha+1)}-\frac{1}{\Gamma^{2}(\alpha+1)}\right)\nonumber\\&+2\bar{\mu}^{2}w_{1}w_{2}t^{\alpha+1}\left(\frac{2}{\Gamma(\alpha+2)}-\frac{1}{\Gamma(\alpha+1)}\right).
\end{align}
Similarly to the behavior of the first log moment, in the log variance, for short time scales the sGBM dynamics dominates. However, we observe that on the long run the dynamics is a combination of the two kernels, since the dominant term is $w_{1}w_{2}t^{\alpha+1}$.

The subordination function for this case is given by
\begin{align}
\hat{h}(u,s)=\frac{1}{w_{1}+w_{2}s^{1-\alpha}}e^{-\frac{u}{w_{1}s^{-1}+w_{2}s^{-\alpha}}},
\end{align}
where the L\'evy exponent is $\hat{\Psi}(s)=\left[w_{1}s^{-1}+w_{2}s^{-\alpha}\right]^{-1}$.

\subsection{Mix of subdiffusive GBMs}\label{sec:gen-gbm.7}

We may further analyze the case of a mix of two sGBM with different power-law memory functions,
\begin{align}\label{N}
\eta(t)=w_{1}\frac{t^{\alpha_{1}-1}}{\Gamma(\alpha_{1})}+w_{2}\frac{t^{\alpha_{2}-1}}{\Gamma(\alpha_{2})},
\end{align}
where $0<\alpha_1<\alpha_2<1$, $w_{1}+w_{2}=1$, and
\begin{align}\label{N Laplace}
\hat{\eta}(s)=w_{1}s^{-\alpha_{1}}+w_{2}s^{-\alpha_{2}}.
\end{align}
This situation corresponds to the case of two different groups of periods of constant prices. For physical systems, this situation means that the particles are trapped in traps with different waiting times \cite{sandev2015distributed}, represented by the memory kernel (\ref{N}).

Therefore, for the mean we find
\begin{align}
\langle x(t)\rangle&=x_0\,\mathscr{L}^{-1}\left\{\frac{s^{-1}}{1-\mu\left(w_{1}s^{-\alpha_1}+w_{2}s^{-\alpha_2}\right)}\right\}(t)=x_0\,\sum_{n=0}^{\infty}w_{1}^{n}\mu^{n}t^{\alpha_1 n}E_{\alpha_2,\alpha_1 n+1}^{n+1}\left(w_{2}\mu t^{\alpha_2}\right), 
\end{align}
while for the MSD we obtain
\begin{align}
\langle x^2(t)\rangle= x_{0}^{2}\,\sum_{n=0}^{\infty}w_{1}^n(\sigma^{2}+2\mu)^{n}t^{\alpha_1 n}E_{\alpha_2,\alpha_1 n+1}^{n+1}\left(w_{2}(\sigma^{2}+2\mu) t^{\alpha_2}\right). 
\end{align} 

Similarly, the log-mean yields
\begin{align}
    \langle \log{x(t)}\rangle=\langle \log{x_0}\rangle+\bar{\mu}\left[w_1\,\frac{t^{\alpha_{1}}}{\Gamma(\alpha_{1}+1)}+w_2\,\frac{t^{\alpha_{2}}}{\Gamma(\alpha_{2}+1)}\right].
\end{align}
The expectation of the log return with period $\Delta{t}$, then becomes 
\begin{align}
    \frac{1}{\Delta{t}}\langle \log{\left(x(t+\Delta{t})/x(t)\right)}\rangle\underset{\Delta{t}\rightarrow0}{\sim}\bar{\mu}\left[w_1\,\frac{t^{\alpha_{1}-1}}{\Gamma(\alpha_{1})}+w_2\,\frac{t^{\alpha_{2}-1}}{\Gamma(\alpha_{2})}\right]=\frac{d}{dt}\langle\log{x(t)}\rangle.
\end{align}
Since $0<\alpha_1<\alpha_2<1$, on short times, the part of first sGBM dominates, whereas on long times it is the characteristic of the second sGBM that determines the dynamics. The second log-moment becomes
\begin{align}
    \langle \log^{2}{x(t)}\rangle&=\langle \log^{2}{x_{0}}\rangle+2\bar{\mu}^{2}\left[\frac{w_{1}^{2}t^{2\alpha_{1}}}{\Gamma(2\alpha_{1}+1)}+\frac{2w_{1}w_{2}t^{\alpha_{1}+\alpha_{2}}}{\Gamma(\alpha_{1}+\alpha_{2}+1)}+\frac{w_{2}^{2}t^{2\alpha_{2}}}{\Gamma(2\alpha_{2}+1)}\right]\nonumber\\&+\left\{2\bar{\mu}\langle \log{x_0}\rangle+\sigma^{2}\right\}\left(w_{1}\frac{t^{\alpha_{1}}}{\Gamma(\alpha_{1}+1)}+w_{2}\frac{t^{\alpha_{2}}}{\Gamma(\alpha_{2}+1)}\right),
\end{align}
and for the log-variance we find \begin{align}
    &\langle \log^{2}{x(t)}\rangle-\langle \log{x(t)}\rangle^{2}=\sigma^{2}\left(w_{1}\frac{t^{\alpha_{1}}}{\Gamma(\alpha_{1}+1)}+w_{2}\frac{t^{\alpha_{2}}}{\Gamma(\alpha_{2}+1)}\right)\nonumber\\&+2\bar{\mu}^{2}w_{1}w_{2}t^{\alpha_{1}+\alpha_{2}}\left(\frac{2}{\Gamma(\alpha_{1}+\alpha_{2}+1)}-\frac{1}{\Gamma(\alpha_{1}+1)\Gamma(\alpha_{2}+1)}\right)\nonumber\\&+\bar{\mu}^{2}w_{1}^{2}t^{2\alpha_{1}}\left(\frac{2}{\Gamma(2\alpha_{1}+1)}-\frac{1}{\Gamma^{2}(\alpha_{1}+1)}\right)+\bar{\mu}^{2}w_{2}^{2}t^{2\alpha_{2}}\left(\frac{2}{\Gamma(2\alpha_{2}+1)}-\frac{1}{\Gamma^{2}(\alpha_{2}+1)}\right).
\end{align}
In this case, for short times, the kernel with the smaller exponent dominates the variance. Interestingly, for long times, this observable is determined by the magnitude of the larger exponent, which is opposite from the previous kernel examples.

For the mix of subdiffusive GBMs the subordination function becomes
\begin{align}
\hat{h}(u,s)=\frac{1}{w_{1}s^{1-\alpha_1}+w_{2}s^{1-\alpha_2}}e^{-\frac{u}{w_{1}s^{-\alpha_1}+w_{2}s^{-\alpha_2}}},
\end{align}
where the L\'evy exponent is $\hat{\Psi}(s)=\left[w_{1}s^{-\alpha_1}+w_{2}s^{-\alpha_2}\right]^{-1}$.

\begin{figure}[!t]
\centering
\includegraphics[width=\linewidth]{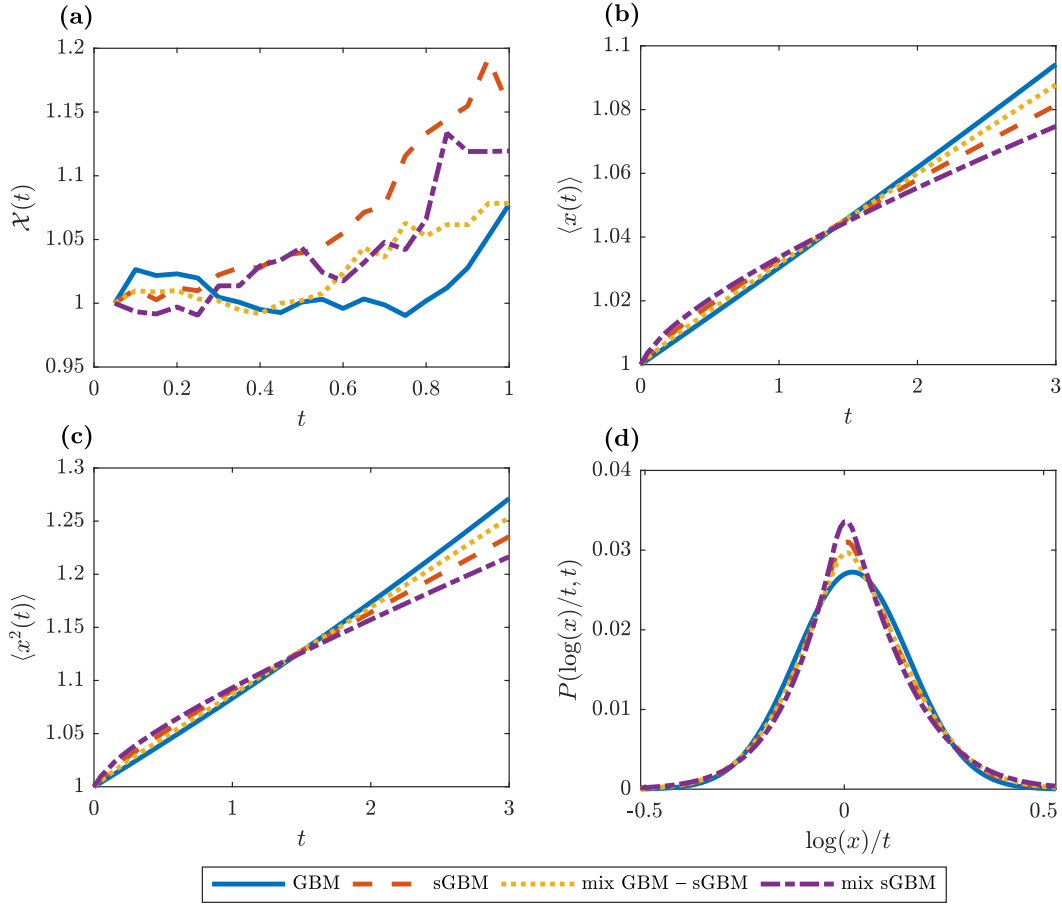}
\caption{\textbf{gGBM properties.} \textbf{(a)}~An example for simulated individual trajectories of gGBM for different memory kernels: standard GBM (blue solid line), sGBM (red dashed line), mix of standard GBM and sGBM (yellow dotted line), mix of sGBM (violet dot-dashed line). \textbf{(b)}~Numerical estimation for the first moment in GBM, sGBM, mix of standard GBM and sGBM and the mix of sGBM as a function of time. \textbf{(b)}~Same as \textbf{(b)}, only for the second moment. \textbf{(c)}~Empirical PDF for the logarithmic return at $t = 1$ estimated from 1000 realizations of gGBM. \textbf{(a)-(c)} In the simulations, $\mu =0.03$ and $\sigma^2= 0.02$. Moreover, for the sGBM case we set $\alpha = 0.8$, for the mix GBM- sGBM case we set $\alpha = 0.8$ and $w_1 = w_2 = 0.5$, and for the mix of sGBM case $\alpha_1 = 0.8$,  $\alpha_2 = 0.6$ and $w_1 = w_2 = 0.5$. \label{fig:memory-kernel-examples}}
\end{figure}

Figure~\ref{fig:memory-kernel-examples}(a), gives an intuitive illustration of the gGBM dynamics under various choices for the kernel. As argued, for standard GBM we observe smooth dynamics without periods of constant prices, whereas there is more turbulence in the asset price dynamics in the gGBM case. The periods of constant prices reproduced by gGBM depend in general on the time scale and, hence, the measuring units of the drift and volatility, with longer time scales also corresponding to longer periods of constant prices. In Figure~\ref{fig:memory-kernel-examples}(b) and Figure~\ref{fig:memory-kernel-examples}(c) we plot, respectively, the numerical approximations for the first moment and the MSD for GBM, sGBM, a mix of GBM and sGBM and a mix of sGBMs. One can easily notice the nonlinear behavior in the generalizations of GBM. For long times all gGBMs give exponential dependence of the first moment and the MSD on time but with smaller slope than the one of GBM. Finally,  Figure~\ref{fig:memory-kernel-examples}(d) gives the empirical PDF for the logarithmic return at $t = 1$. For each of the studied generalizations of GBM, the PDF is characterised with fatter tails (which should increase as the $\alpha$ parameters increase), meaning that it is more prone to producing values that fall far from the average. This can be easily observed as from the excess kurtosis present in each GBM generalization. This is exactly what makes the gGBM framework useful for understanding the statistical behavior of the asset price dynamics.

\section{Empirical example}\label{sec:empirical-example}

To illustrate the power of the gGBM framework in the description of option pricing we utilise empirical data of American options for two companies, Tesla (TSLA) and Apple (AAPL). By definition, the dynamics of American options differ from European as they allow exercising of the option at any time before the option expires. Nevertheless, as given in Hull (2017), one can rely on the fact that American options on non-dividend-paying stocks have the same value as their European counterpart. This relation has allowed for the empirical examination of a pricing scheme of European options to be widely done via data for American ones.

For our analysis we use the freely available data from the Nasdaq’s Options Trading Center. This dataset offers daily data free of charge for options of all companies quoted there. However, the options for most companies have small sample size. Therefore we have restricted the empirical analysis to Tesla and Apple, whose options are more frequently traded. In our estimations, the drift parameter $\mu$ is simply taken as the 3-Month Treasury Bill Secondary Market Rate at the date of observation. The noise parameter, on the other hand, was  inferred from the values of the options on the market as the value which produces the minimum squared error in their fit. In finance, this is known as use the famous ``implied volatility'' approach.

\begin{figure}[!t]
 \includegraphics[width=\linewidth]{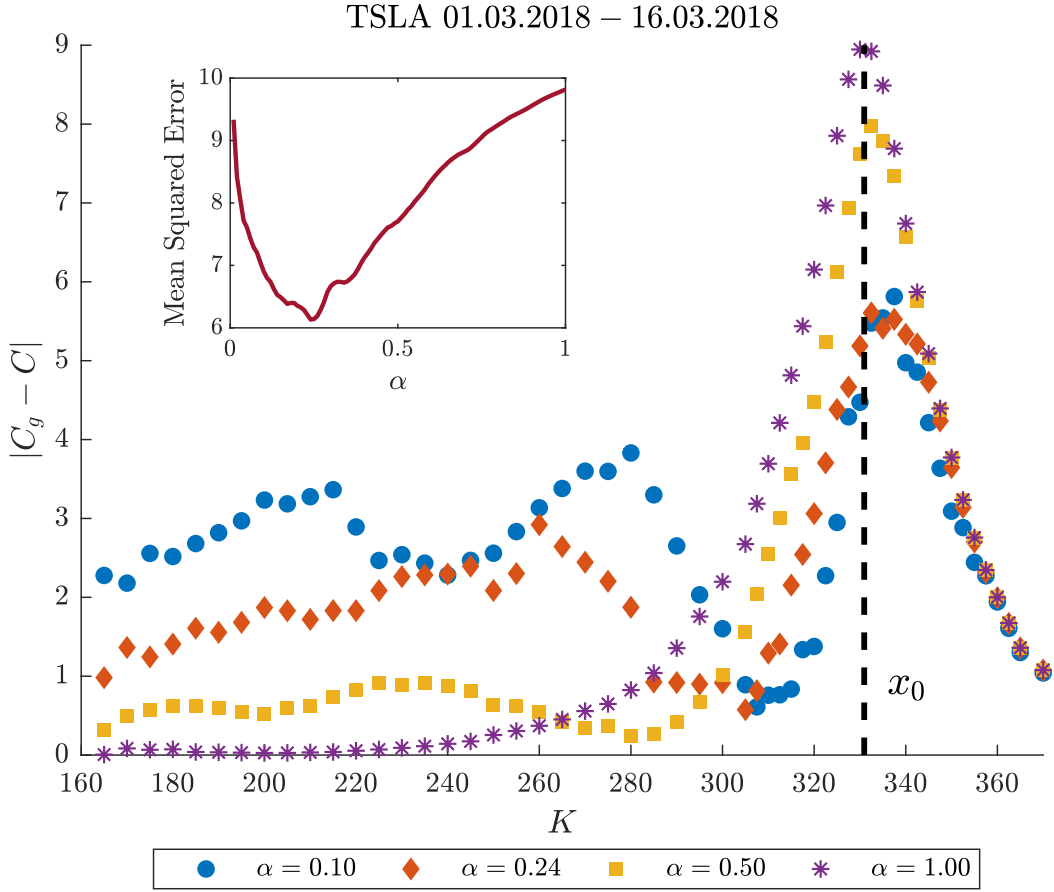}
 \caption{\textbf{Moneyness in sGBM.}  The absolute difference between the predicted TSLA option price $C_g$ and its real value $C$ as a function of the strike price of the option for various choices of $\alpha$. The data is taken on 1st March 2020 and describe the value of TSLA options which expire on 16th March 2020. The inset plot gives the mean squared error of the predictions as a function of $\alpha$.}
\label{fig:moneyness}
\end{figure}

Let us now turn our attention to Fig.~\ref{fig:moneyness} where we use TSLA data gathered on 1st March 2018 on options which expire on 16th March 2018 to examine the dependence of the sGBM model on the moneyness of the option in predicting it. Moneyness describes the relative position of the current price of TSLA ($x_0$) with respect to the strike price of the option. An option whose strike price is equal to the current price of the asset is said to be at the money; if the strike price is larger than the current price, the option is ``out of the money''; and if the strike price is smaller than the current price, the option is described to be ``in the money''. In Fig.~\ref{fig:moneyness} we vary the subdiffusion parameter $\alpha$, and plot the absolute difference in the estimated option price $C_g$ and the observed option price as a function of the strike price. We find that for in-the-money-options the best prediction is with $\alpha = 1$, which corresponds to the BS model. However, as the strike price of the option nears the TSLA price, a transition occurs and $\alpha = 1$ becomes the worst predictor of the option price, whereas the lower the subdiffusion parameter, the better prediction we get. For options that are out of the money, it appears that the performance of the prediction for the option price does not depend on $\alpha$. Overall, as shown in the inset plot where we plot the mean squared error of the prediction as a function of $\alpha$, this analysis suggests that the best prediction for the TSLA data is done with $\alpha$ which is around $0.25$, thus highlighting the subdiffusive nature of the dynamics of the TSLA stock.

\begin{figure}[!t]
 \includegraphics[width=\linewidth]{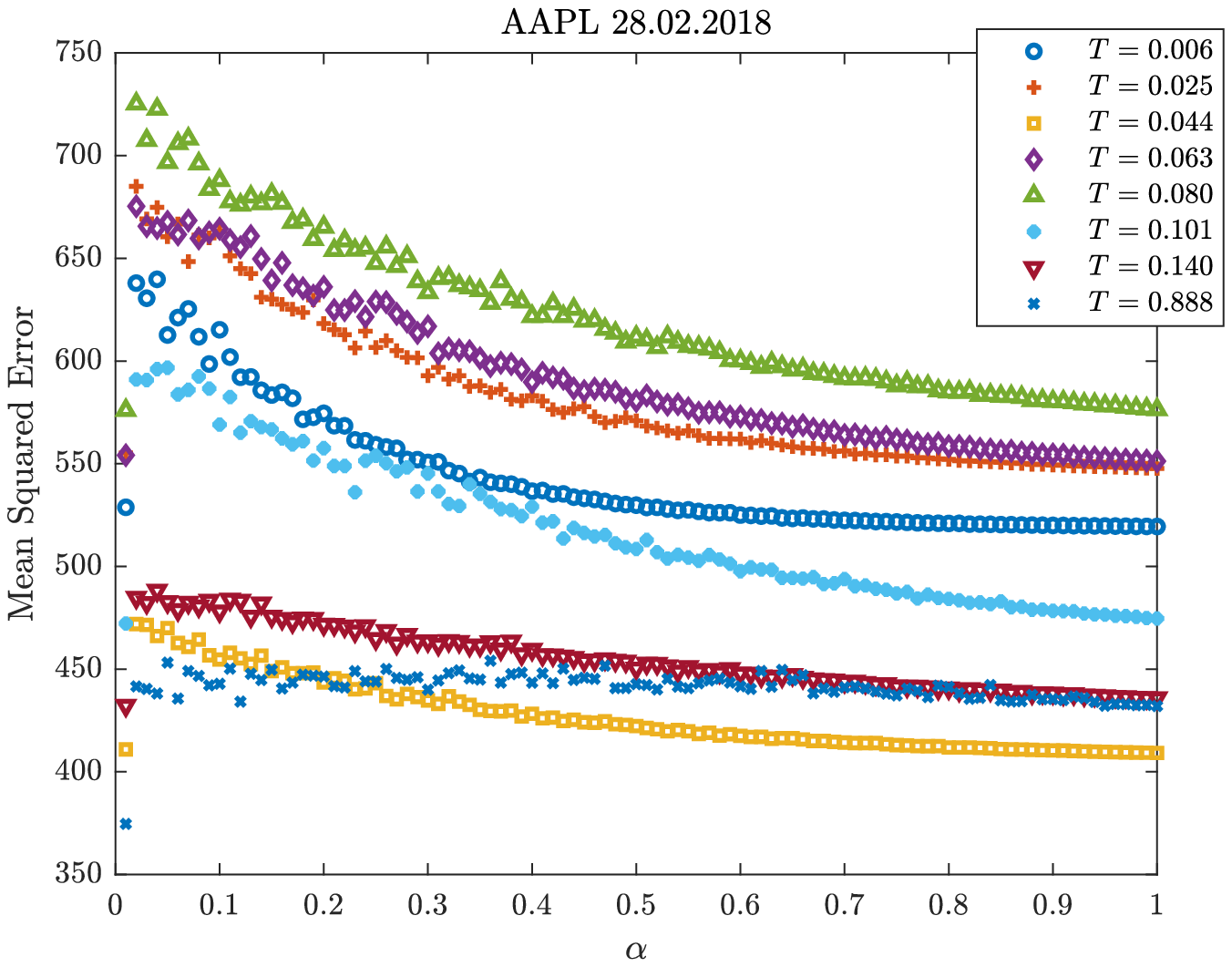}
 \caption{\textbf{Maturity in sGBM.} Mean squared error of the prediction of the AAPL option price with data taken on 28th February 2018 as a function of $\alpha$ for various maturity periods $T$ (measured in years).}
\label{fig:maturity}
\end{figure}

Next, we use the AAPL data gathered on 28th February 2018 and examine how the maturity $T$ affects the performance of the same sGBM model in predicting the option price. For this purpose, Fig.~\ref{fig:maturity} depicts the mean squared error of the option price prediction as a function of the parameter $\alpha$. We observe that, in general, the best prediction occurs when $\alpha = 1$. This may suggest that the dynamics of the AAPL stock price is quite nicely explained with the BS model. However, we also see that the mean squared error is highly dependent on the maturity, and even that for some maturity very low subdiffusive parameter values exhibit similar performance as the BS model. Hence, one might even argue that different gGBM kernels can lead to similar outcomes in the pricing of options, an interesting finding as such.

Evidently, the performance of a kernel ultimately depends on the  physical properties of the option. On the first sight, this conclusion appears intuitive -- obviously the known information for the properties of the asset greatly impacts its price, the observation that a slight change in the known information may drastically change the dynamics suggests that there is a need in the option pricing literature for models that easily allow for such structural changes. In this aspect, we believe that the generalised GBM approach offers a computationally inexpensive and efficiently tractable solution to this issue. Consequently, we stress that a significant improvement of the description of the data in the gGBM framework can be achieved with comparatively few additional parameters.


\section{Conclusion}\label{sec:conclusion}

We investigated the potential of GBM extensions based on subdiffusion to model and predict the price of options. By assuming that the price of the asset underlying the option undergoes a subdiffusive process, we introduced the gGBM framework as a potential model for its value.

Similar to previous works on subdiffusive GBM models, the dynamics of a particular gGBM instance is critically determined by a memory kernel. The advantage of gGBM comes in the flavor of allowing various forms for the functional form of the kernel. Depending on its choice, we may end up with asset price dynamics whose behavior significantly varies on the short time in comparison to its long run characteristics. This, in turn, may induce observations of the properties of the asset price that more closely mimic realistic behavior than standard GBM.

We explored the ability of gGBM to fit and predict real option values. Our empirical analysis confirmed the characteristics of gGBM, as we discovered that the performance of a certain choice of memory kernel is uniquely determined by the parameters of the option, such as its maturity and its moneyness. Since each kernel produces, in general, different long run and short run dynamics, this suggests that time-averages play an important role in efficient pricing of options. Formally, time-averaging is essential for the analysis of a single time-series (or a set of few) which is characterised with non-ergodic dynamics. The non-ergodicity creates non-equilibrium dynamics which, consequently, makes studies of the ensemble behavior irrelevant. This leads to the introduction of novel strategies for analysing financial data~\cite{1,cherstvy}.

In line with our conclusions, we believe that the next step in uncovering the properties of gGBM is demonstrating the ergodicity breaking of the process. Since multiplicative processes are frequently present in nature, this will not only extend the framework of gGBM in analysing financial data, but will also provide an avenue for applying the model in other scientific domains. Another fruitful research direction would be to incorporate the properties of gGBM in a wider framework for financial modeling which includes the concept of ``rough volatility'', where the instantaneous volatility is driven by a (rough) fractional Brownian motion~\cite{el2018quantitative}.  Building an explanatory model for the volatility in terms of gGBM would bring novel insights about the theoretical and empirical characteristics of the asset prices. We also leave for future analysis the problem of gGBM with stochastic volatility, which can be treated in the framework of the Fokker-Planck equation for gGBM with time varying volatility $\sigma(t)$, in analogy of the diffusing-diffusivity models for heterogeneous media \cite{dd1,dd2,dd3,dd4,dd5}. 

\vspace{6pt} 

\section*{Ackdnowledgments}

The Authors acknowledge funding from the Deutsche Forschungsgemeinschaft (DFG). TS was supported by the Alexander von Humboldt Foundation. TS acknowledges Dr. Andrey Cherstvy for the fruitful discussions and suggestions.



\section*{Abbreviations}

The following abbreviations are used in this manuscript:

\noindent 
\begin{tabular}{@{}ll}
GBM & Geometric Brownian motion\\
sGBM & Subdiffusive geometric Brownian motion\\
gGBM & Generalised geometric Brownian motion\\
BS & Black-Scholes\\
CTRW & Continuous time random walk\\
MSD & Mean squared displacement\\
ML & Mittag-Leffler\\
TSLA & Tesla\\
AAPL & Apple
\end{tabular}


\appendix

\section{Solution of the Fokker-Planck equation for standard GBM}\label{app1}

\setcounter{equation}{0}
\setcounter{figure}{0}
\setcounter{table}{0}
\setcounter{theorem}{0}
\makeatletter
\renewcommand{\theequation}{A\arabic{equation}}
\renewcommand{\thetable}{A\arabic{table}}
\renewcommand{\thefigure}{A\arabic{figure}}
\renewcommand{\thetheorem}{A\arabic{theorem}}
\renewcommand{\theproposition}{A\arabic{proposition}}

The solution of Eq.~(\ref{generalizedFPE_standard}) can be found by using the Laplace-Mellin transform method \cite{submitted}. The Laplace transform is defined by $$\hat{F}(s)=\mathscr{L}\left\{f(t)\right\}(s)=\int_{0}^{\infty}f(t)\,e^{-st}\,dt,$$ while the Mellin transform as \cite{8} $$\tilde{F}(q)=\mathscr{M}\left\{f(x)\right\}(q)=\int_{0}^{\infty}x^{q-1}\,f(x)\,dx.$$ The inverse Mellin transform then reads $$f(x)=\mathscr{M}^{-1}\left\{\tilde{F}(q)\right\}(x)=\frac{1}{2\pi\imath}\int_{c-\imath\infty}^{c+\imath\infty}x^{-q}\,\tilde{F}(q)\,dq.$$ Therefore, by performing Laplace transform in respect to $t$ and Mellin transform in respect to $x$ in Eq.~(\ref{generalizedFPE_standard}), we have
\begin{align}\label{generalizedFPE_standard_app mellin}
\tilde{\hat{F}}(q,s)=x_{0}^{q-1}\times\frac{1}{s-\left[\frac{\sigma^2}{2}(q-1)(q-2)+\mu(q-1)\right]},
\end{align}
where we use $\mathscr{M}\left\{\delta(x-x_0)\right\}(q)=x_{0}^{q-1}$. Then the inverse Laplace transform yields
\begin{align}
    \tilde{F}(q,t)&=x_{0}^{q-1}\times\exp\left(\frac{\sigma^2}{2}\left[q+\frac{1}{2}\left(\frac{2\mu}{\sigma^{2}}-3\right)\right]^2t-\frac{\left(\mu-\frac{\sigma^2}{2}\right)^2}{2\sigma^2}t\right),
\end{align}
where we use $\mathscr{L}^{-1}\left\{\frac{1}{s-a}\right\}(t)=e^{at}$. Applying the inverse Mellin transform and looking for the solution in the form of the convolution integral of two functions~\cite{8}, $\mathscr{M}\left\{h(x)\right\}(q)=\tilde{H}(q)$ and $\mathscr{M}\left\{g(x)\right\}(q)=\tilde{G}(q)$, $$\mathscr{M}^{-1}\left\{\tilde{H}(q)\,\tilde{G}(q)\right\}(x)=\int_{0}^{\infty}h(r)\,g(x/r)\frac{dr}{r},$$we obtain the solution of the Fokker-Planck equation for GBM
\begin{align}\label{solution_GFP}
    f(x,t)&=\int_{0}^{\infty}\delta(r-x_0)\times\frac{\exp\left(-\frac{\left[\log\frac{x}{r}-\left(\mu-\frac{\sigma^2}{2}\right) t\right]^{2}}{2\sigma^{2}t}\right)}{(x/r) \sqrt{2\pi\sigma^{2}t}}\frac{dr}{r}\nonumber\\
    &=\frac{1}{x\sqrt{2\pi\sigma^2t}}\times\exp\left(-\frac{\left[\log\frac{x}{x_0}-\left(\mu-\frac{\sigma^2}{2}\right) t\right]^{2}}{2\sigma^{2}t}\right).
\end{align}
Here we use that $$h(x)=\mathscr{M}^{-1}\left\{x_{0}^{q-1}\right\}(x)=\delta(x-x_0)$$and
\begin{align}
    g(x)&=\mathscr{M}^{-1}\left\{\exp\left(\frac{\sigma^2}{2}\left[q+\frac{1}{2}\left(\frac{2\mu}{\sigma^{2}}-3\right)\right]^2t-\frac{\left(\mu-\frac{\sigma^2}{2}\right)^2}{2\sigma^2}t\right)\right\}(x)\nonumber\\&=\frac{1}{x\sqrt{2\pi\sigma^2t}}\times\exp\left(-\frac{\left[\log{x}-\left(\mu-\frac{\sigma^2}{2}\right) t\right]^{2}}{2\sigma^{2}t}\right).\nonumber
\end{align}
We also used the properties of the inverse Mellin transform \cite{8}, $\mathscr{M}^{-1}\left\{f(q+a)\right\}(x)=x^{a}\mathscr{M}^{-1}\left\{f(q)\right\}$ and $$\mathscr{M}^{-1}\left\{\exp\left(\alpha q^2\right)\right\}(x)=\frac{1}{\sqrt{4\pi\alpha}}\,e^{-\frac{x^2}{4\alpha}}.$$
Therefore, from the solution~(\ref{solution_GFP}) we conclude that the solution of the Fokker-Planck equation is a log-normal distribution. 

The $n$th moment $\langle x^{n}(t)\rangle=\int_{0}^{\infty}x^{n}P(x,t)\,dx$ of the solution of Eq.~(\ref{generalizedFPE_standard}) can be obtained by multiplying the both sides of the equation with $x^n$ and integration over $x$. Thus, one has
\begin{align}\label{n-th moment gbm}
\frac{\partial}{\partial t}\langle x^{n}(t)\rangle=\left[\frac{\sigma^{2}}{2}n(n-1)+\mu\,n\right]\langle x^{n}(t)\rangle,
\end{align}
from where the $n$th moment becomes
\begin{align}
    \langle x^{n}(t)\rangle=\langle x^{n}(0)\rangle\,e^{\left(\sigma^{2}n(n-1)/2+\mu\,n\right)t}.
\end{align}
For $n=0$ one observes that the solution of the Fokker-Planck equation for GBM is normalised, i.e., $\langle x^{0}(t)\rangle=1$. The mean value ($n=1$) and the MSD have exponential dependence on time, $\langle x(t)\rangle=\langle x(0)\rangle\,e^{\mu\,t}$ and $\langle x^{2}(t)\rangle=\langle x^{2}(0)\rangle\,e^{(\sigma^{2}+2\,\mu)t}$, respectively, and thus, the variance becomes
\begin{align}
    \langle x^{2}(t)\rangle-\left\langle x(t)\right\rangle^{2}= \langle x^{2}(0)\rangle\,e^{2\mu t}\left(e^{\sigma^{2}t}-1\right).
\end{align}

The log-moments $\langle\log^{n}{x}\rangle=\int_{0}^{\infty}\log^{n}{x}P(x,t)\,dx$ can be obtained by multiplying the both sides of Eq.~(\ref{generalizedFPE_standard}) with $\log^{n}x$ and integration over $x$. Therefore, one finds the following equation (see Ref.~\cite{20} for details)
\begin{align}
    \frac{\partial}{\partial t}\langle \log^{n}{x(t)}\rangle=\left(\mu-\frac{\sigma^{2}}{2}\right)n\,\langle \log^{n-1}{x(t)}\rangle+\frac{\sigma^{2}}{2}n(n-1)\langle \log^{n-2}{x(t)}\rangle.
\end{align}
From here it follows that $\frac{\partial}{\partial t}\langle \log^{0}{x(t)}\rangle=0$, i.e., $\langle \log^{0}{x(t)}\rangle=\langle \log^{0}{x(0)}\rangle=1$. The case $n=1$ yields the mean value of the logarithm of $x(t)$,
\begin{align}
    \frac{\partial}{\partial t}\langle \log{x(t)}\rangle=\left(\mu-\frac{\sigma^{2}}{2}\right)\underbrace{\langle \log^{0}{x(t)}\rangle}_{=1}
\end{align}
i.e.,
\begin{align}
    \langle \log{x(t)}\rangle=\langle \log{x(0)}\rangle+\left(\mu-\frac{\sigma^{2}}{2}\right)t.
\end{align}
For $n=2$ we obtain the second log-moment
\begin{align}
    &\frac{\partial}{\partial t}\langle \log^{2}{x(t)}\rangle=2\left(\mu-\frac{\sigma^{2}}{2}\right)\langle \log{x(t)}\rangle+\sigma^{2}\langle \log^{0}{x(t)}\rangle 
\end{align}
which is given by
\begin{align}
    \langle \log^{2}{x(t)}\rangle=\langle \log^{2}{x(0)}\rangle+\left(\mu-\frac{\sigma^{2}}{2}\right)^{2}t^{2}+2\left(\mu-\frac{\sigma^{2}}{2}\right)\langle \log{x(0)}\rangle t+\sigma^{2}t.
\end{align}
Therefore, for the log-variance one finds linear dependence on time
\begin{align}
    \langle \log^{2}{x(t)}\rangle-\langle \log{x(t)}\rangle^{2}=\sigma^{2}t.
\end{align}

\section{Derivation of the Fokker-Planck equation for gGBM from CTRW theory}\label{sec:gen-gbm.2}

\setcounter{equation}{0}
\setcounter{figure}{0}
\setcounter{table}{0}
\setcounter{theorem}{0}
\makeatletter
\renewcommand{\theequation}{B\arabic{equation}}
\renewcommand{\thetable}{B\arabic{table}}
\renewcommand{\thefigure}{B\arabic{figure}}
\renewcommand{\thetheorem}{B\arabic{theorem}}
\renewcommand{\theproposition}{B\arabic{proposition}}

We use the approach given in Refs.~\cite{18,18_2,18_3}. Let us consider a CTRW for a particle at position $x_{i}$ which can move right to the position $x_{i+1}=h\,x_{i}$ or to left at position $x_{i-1}=\frac{1}{h}\,x_{i}$, $h>0$. For the CTRW on a geometric lattice we use $h=1+u$, and at the end we will find the diffusion limit $u\rightarrow0$. The probability density function (PDF) for the particle to jump to right is $p_{\text{r}}(x,t)$, and for jump to left $p_{\textrm{l}}(x,t)$. The total probability is $p_{\text{r}}(x,t)+p_{\textrm{l}}(x,t)=1$.

We consider a multiplicative jump length PDF on a geometric lattice \cite{18}, $$\lambda(x_{i},t,x_{j})=p_{\textrm{r}}(x_{j},t)\,\delta(x_{i}-[1+u]\,x_{j})+p_{\textrm{l}}(x_{j},t)\,\delta(x_{i}-x_{j}/[1+u]),$$and a waiting time PDF $\psi(t)$, related to the survival probability by $$\phi(t)=1-\int_{0}^{t}\psi(t')\,dt', \quad \textrm{i.e.,} \quad \hat{\phi}(s)=\frac{1-\hat{\psi}(s)}{s}.$$ By substitution in the master equation \cite{18} 
\begin{align}
    \frac{\partial}{\partial t}\rho(x_{i},t)=\sum_{j}\lambda(x_{i},t,x_{j})\int_{0}^{t}K(t-t')\,\rho(x_{j},t')\,dt'-\int_{0}^{t}K(t-t')\,\rho(x_{i},t')\,dt',
\end{align}
where $K(t)=\mathscr{L}^{-1}\left[\hat{\psi}(s)/\hat{\phi}(s)\right]$, one finds
\begin{align}\label{master K}
    \frac{\partial}{\partial t}\rho(x_{i},t)&=p_{\textrm{r}}\left(\frac{x_{i}}{1+u},t\right)\int_{0}^{t}K(t-t')\,\rho\left(\frac{x_{i}}{1+u},t'\right)\,dt'\nonumber\\&+p_{\textrm{l}}\left([1+u]x_{i},t\right)\int_{0}^{t}K(t-t')\,\rho\left([1+u]x_{i},t'\right)\,dt'-\int_{0}^{t}K(t-t')\,\rho(x_{i},t')\,dt'.
\end{align}
We consider generalised waiting time PDF, which in the Laplace space has the form \cite{19,19_2} $$\hat{\psi}(s)=\frac{1}{1+\tau_{\eta}/\hat{\eta}(s)},$$where $\tau_{\eta}$ is a time parameter, which depends on $\eta(t)$. Therefore, $$\hat{\phi}(s)=\frac{\tau_{\eta}/\hat{\eta}(s)}{s\left(1+\tau_{\eta}/\hat{\eta}(s)\right)},$$and $$\hat{K}(s)=\frac{1}{\tau_{\eta}}\,s\times\hat{\eta}(s),$$from where we find that $\int_{0}^{t}K(t-t')\,f(t')\,dt'\rightarrow\frac{1}{\tau_{\eta}}\frac{d}{dt}\int_{0}^{t}\eta(t-t')\,f(t')\,dt'$. From Eq.~(\ref{master K}) then we obtain
\begin{align}\label{master eta}
    \frac{\partial}{\partial t}\rho(x_{i},t)&=\frac{1}{\tau_{\eta}}\,p_{\textrm{r}}\left(\frac{x_{i}}{1+u},t'\right)\frac{\partial}{\partial t}\int_{0}^{t}\eta(t-t')\,\rho\left(\frac{x_{i}}{1+u},t'\right)\,dt'\nonumber\\&+\frac{1}{\tau_{\eta}}\,p_{\textrm{l}}\left([1+u]x_{i},t'\right)\frac{\partial}{\partial t}\int_{0}^{t}\eta(t-t')\,\rho\left([1+u]x_{i},t'\right)\,dt'-\frac{1}{\tau_{\eta}}\,\frac{\partial}{\partial t}\int_{0}^{t}\eta(t-t')\,\rho(x_{i},t')\,dt'.
\end{align}
Let us now consider the diffusion limit ($u\rightarrow0$ and $\tau_{\eta}\rightarrow0$) of Eq.~(\ref{master eta}). From the normalisation condition of the PDF $\rho(x,t)$ given by $\sum_{i}\rho(x_i,t)=1$ and by using position-dependent lattice spacing $\Delta x_{i}=u\,x_{i}$, one finds
$\sum_{i}\frac{\rho_{u}(x_i,t)}{u\,x_i}\Delta x_{i}=1$, such that $\lim_{\Delta x_i\rightarrow0}\sum_{i}\left(\frac{\rho_{u}(x_i,t)}{u\,x_i}\right)\Delta x_{i}=1$ \cite{18}. By defining the function $P_{u}(x,t)=\rho_{u}(x,t)/[u\,x]$, one concludes that $P(x,t)=\lim_{u\rightarrow0}P_{u}(x,t)$ is normalised, i.e., $\int_{0}^{\infty}P(x,t)\,dx=1$. By introducing $B_{u}(x_{i},t)=p_{\textrm{r}}(x_{i},t)-p_{\textrm{l}}(x_{i},t)$ and $b_{0}(x,t)=\lim_{u\rightarrow0}\frac{\partial}{\partial u}B_{u}(x,t)$, in the diffusion limit $u\rightarrow0$ and $\tau_{\eta}\rightarrow0$, where we assume that $B_{0}(x,t)=\lim_{u\rightarrow0}B_{u}(x,t)=0$ \cite{18}, we arrive to the following Fokker-Planck equation
\begin{align}
    \frac{\partial}{\partial t}P(x,t)=\mathcal{D}\frac{\partial}{\partial t}\int_{0}^{t}\eta(t-t')\frac{\partial}{\partial x}\left(x^{2}\frac{\partial}{\partial x}-\frac{1}{k_{B}T}x^{2}\,F(x)\right)P(x,t')\,dt',
\end{align}
where $\mathcal{D}=\lim_{u,\tau_{\eta}\rightarrow0}u^{2}/[2\,\tau_{\eta}]$, $F(x)=-V'(x)=k_{B}T\,[2\,b_{0}(x)-1]/x$, and $P(x,t)\propto \exp\left(-\frac{V(x)}{k_{B}T}\right)$ is obtained from the long time steady state Boltzmann distribution \cite{18}. For a logarithmic potential $V(x)=v\,k_{B}T\,\log{x}$, the force becomes $F(x)=-k_{B}T\,v/x$. By using $\mathcal{D}=\sigma^{2}/2$ and $v=2-\mu/\mathcal{D}$ the Fokker-Planck equation becomes
\begin{align}
    \frac{\partial}{\partial t}P(x,t)=\frac{\partial}{\partial t}\int_{0}^{t}\eta(t-t')\frac{\partial}{\partial x}\left(\frac{\sigma^{2}x^{2}}{2}\frac{\partial}{\partial x}+[\sigma^{2}-\mu] x\right)P(x,t')\,dt',
\end{align}
which can be rewritten in the form of Eq.~(\ref{generalizedFPE2}).

\section{General results for \textit{n}th moment}\label{app3}

\setcounter{equation}{0}
\setcounter{figure}{0}
\setcounter{table}{0}
\setcounter{theorem}{0}
\makeatletter
\renewcommand{\theequation}{C\arabic{equation}}
\renewcommand{\thetable}{C\arabic{table}}
\renewcommand{\thefigure}{C\arabic{figure}}
\renewcommand{\thetheorem}{C\arabic{theorem}}
\renewcommand{\theproposition}{C\arabic{proposition}}

If we multiply both sides of Eq.~(\ref{generalizedFPE2}) by $x^n$, and integrate over $x$ we find the $n$th moment $\langle x^{n}(t)\rangle=\int_{0}^{\infty}x^{n}P(x,t)\,dx$,
\begin{align}\label{n-th moment}
\frac{\partial}{\partial t}\langle x^{n}(t)\rangle=\left[\frac{\sigma^{2}}{2}n(n-1)+\mu\,n\right]\mu\frac{d}{dt}\int_{0}^{t}\eta(t-t')\langle x^{n}(t')\rangle\,dt',
\end{align}
from where in the Laplace space it reads
\begin{align}\label{n-th moment laplace app}
\langle \hat{x}^{n}(s)\rangle=\frac{s^{-1}}{1-\hat{\eta}(s)\left[\frac{\sigma^{2}}{2}n(n-1)+\mu\,n\right]}\langle x^{n}(0)\rangle.
\end{align}
From this result we obtain the normalization condition, $\frac{\partial}{\partial t}\langle x^{0}(t)\rangle=0$, i.e., $\langle x^{0}(t)\rangle=\langle x^{0}(0)\rangle=1$. For $n=1$, we find the equation for the mean value
\begin{align}
\frac{\partial}{\partial t}\left\langle x(t)\right\rangle=\mu\frac{d}{dt}\int_{0}^{t}\eta(t-t')\left\langle x(t')\right\rangle\,dt',
\end{align}
and its Laplace pair
\begin{align}\label{mean general eta app}
\left\langle \hat{x}(s)\right\rangle=\frac{s^{-1}}{1-\mu\hat{\eta}(s)}\left\langle x(0)\right\rangle.
\end{align}
In terms of the memory kernel $\gamma(t)$, Eq.~(\ref{mean general eta app}) reads
\begin{align}\label{mean general gamma}
\left\langle \hat{x}(s)\right\rangle=\frac{\hat{\gamma}(s)}{s\hat{\gamma}(s)-\mu}\left\langle x(0)\right\rangle.
\end{align}
We note that for the standard case with $\eta(t)=1$ ($\hat{\eta}(s)=1/s$) we recover the previously obtained results for the GBM. For $n=2$ we obtain the equation for the second moment, or the MSD,
\begin{align}
\frac{\partial}{\partial t}\langle x^2(t)\rangle=(\sigma^{2}+2\mu)\frac{d}{dt}\int_{0}^{t}\eta(t-t')\langle x^2(t')\rangle\,dt',
\end{align}
and its Laplace pair
\begin{align}
\langle \hat{x}^2(s)\rangle=\frac{s^{-1}}{1-(\sigma^{2}+2\mu)\hat{\eta}(s)}\langle x^2(0)\rangle,
\end{align}
or
\begin{align}
\langle \hat{x}^2(s)\rangle=\frac{\hat{\gamma}(s)}{s\hat{\gamma}(s)-(\sigma^{2}+2\mu)}\langle x^2(0)\rangle.
\end{align}

We also calculate the log-moments $\left\langle \log^{n}{x(t)}\right\rangle=\int_{0}^{\infty}\log^{n}x\,P(x,t)\,dx$, which satisfy the following integral equation
\begin{align}
    \frac{\partial}{\partial t}\langle \log^{n}{x(t)}\rangle=\frac{\partial}{\partial t}\int_{0}^{t}\eta(t-t')\left[\left(\mu-\frac{\sigma^{2}}{2}\right)n\,\langle \log^{n-1}{x(t')}\rangle+\frac{\sigma^{2}}{2}n(n-1)\,\langle \log^{n-2}{x(t')}\rangle\right]dt'.
\end{align}
Thus, we find that $\frac{\partial}{\partial t}\langle \log^{0}{x(t)}\rangle=0$, i.e., $\langle \log^{0}{x(t)}\rangle=\langle \log^{0}{x(0)}\rangle=1$. For the mean value ($n=1$), we find
\begin{align}
    &\frac{\partial}{\partial t}\langle \log{x(t)}\rangle=\left(\mu-\frac{\sigma^{2}}{2}\right)\frac{\partial}{\partial t}\int_{0}^{t}\eta(t-t')\underbrace{\langle \log^{0}{x(t')}\rangle}_{=1}\,dt' 
\end{align}
from where it follows
\begin{align}
    \langle \log{x(t)}\rangle=\langle \log{x(0)}\rangle+\left(\mu-\frac{\sigma^{2}}{2}\right)\int_{0}^{t}\eta(t')\,dt'.
\end{align}
For the expectation of the periodic log return with period $\Delta{t}$, we find
\begin{align}
    \frac{1}{\Delta{t}}\langle \log{\left(x(t+\Delta{t})/x(t)\right)}\rangle&=\left(\mu-\frac{\sigma^{2}}{2}\right)\frac{1}{\Delta{t}}\int_{t}^{t+\Delta{t}}\eta(t')\,dt'\nonumber\\&=\left(\mu-\frac{\sigma^{2}}{2}\right)\frac{I(t+\Delta t)-I(t)}{\Delta t}\underset{\Delta{t}\rightarrow0}{\sim}\left(\mu-\frac{\sigma^{2}}{2}\right)\eta(t),
\end{align}
where $I(t)=\int\eta(t)\,dt$, i.e., $I'(t)=\eta(t)$. Therefore, the expectation of the periodic log returns behaves as the rate of the first log-moment,
\begin{align}
    \frac{1}{\Delta{t}}\langle \log{\left(x(t+\Delta{t})/x(t)\right)}\rangle\underset{\Delta{t}\rightarrow0}{\sim}\frac{d}{dt}\langle\log{x(t)}\rangle.
\end{align}
For $n=2$ we obtain the second log-moment
\begin{align}
    &\frac{\partial}{\partial t}\langle \log^{2}{x(t)}\rangle=\frac{\partial}{\partial t}\int_{0}^{t}\eta(t-t')\left[2\left(\mu-\frac{\sigma^{2}}{2}\right)\langle \log{x(t')}\rangle+\sigma^{2}\,\langle \log^{0}{x(t')}\rangle\right]dt' 
\end{align}
i.e.,
\begin{align}
    \langle \log^{2}{x(t)}\rangle&=\langle \log^{2}{x(0)}\rangle\nonumber\\&+\int_{0}^{t}\eta(t-t')\left\{2\left(\mu-\frac{\sigma^{2}}{2}\right)\left[\langle \log{x(0)}\rangle+\left(\mu-\frac{\sigma^{2}}{2}\right)\int_{0}^{t'}\eta(t'')\,dt''\right]+\sigma^{2}\right\}dt',
\end{align}
and the log-variance becomes
\begin{align}
    &\langle \log^{2}{x(t)}\rangle-\langle \log{x(t)}\rangle^{2}\nonumber\\&=\sigma^{2}\int_{0}^{t}\eta(t')\,dt'+\left(\mu-\frac{\sigma^{2}}{2}\right)^{2}\left[2\int_{0}^{t}\eta(t-t')\left(\int_{0}^{t'}\eta(t'')\,dt''\right)dt'-\left(\int_{0}^{t}\eta(t')\,dt'\right)^2\right].
\end{align}

\section{Fox {\it H}-function}\label{appH}

\setcounter{equation}{0}
\setcounter{figure}{0}
\setcounter{table}{0}
\setcounter{theorem}{0}
\makeatletter
\renewcommand{\theequation}{D\arabic{equation}}
\renewcommand{\thetable}{D\arabic{table}}
\renewcommand{\thefigure}{D\arabic{figure}}
\renewcommand{\thetheorem}{D\arabic{theorem}}
\renewcommand{\theproposition}{D\arabic{proposition}}

The Fox $H$-function is defined by~\cite{saxena_book}
\begin{eqnarray}\label{H_integral}
H_{p,q}^{m,n}\left[z\left|\begin{array}{l l}
    (a_1,A_1),\dots,(a_p,A_p)\\
    (b_1,B_1),\dots,(b_q,B_q)
  \end{array}\right.\right]=H_{p,q}^{m,n}\left[z\left|\begin{array}{l l}
    (a_p,A_p)\\
    (b_q,B_q)
  \end{array}\right.\right]=\frac{1}{2\pi\imath}\int_{\Omega}\theta(s)z^{s}\,ds,\nonumber\\
\end{eqnarray}
where $\theta(s)$ is given by
$\theta(s)=\frac{\prod_{j=1}^{m}\Gamma(b_j-B_js)\prod_{j=1}^{n}\Gamma(1-a_j+A_js)}{\prod_{j=m+1}^{q}\Gamma(1-b_j+B_js)\prod_{j=n+1}^{p}\Gamma(a_j-A_js)}$, $0\leq n\leq p$, $1\leq m\leq q$, $a_i,b_j \in C$, $A_i,B_j \in R^{+}$, $i=1,...,p$, $j=1,...,q$. The contour $\Omega$ starting at $c-\imath\infty$ and ending at $c+\imath\infty$ separates the poles of the function $\Gamma(b_j+B_js)$, $j=1,...,m$ from those of the function $\Gamma(1-a_i-A_is)$, $i=1,...,n$. A special case of the Fox $H$-function is the exponential function~\cite{saxena_book},
\begin{align}\label{exp vs h}
    e^{-z}=H_{0,1}^{1,0}\left[z\left|\begin{array}{l l}
    - \\
    (0,1)
  \end{array}\right.\right].
\end{align}
The inverse Laplace transform of the Fox $H$-function reads~\cite{saxena_book}
\begin{eqnarray}\label{h laplace}
    \mathscr{L}^{-1}\left[s^{-\rho}\,H_{p,q}^{m,n}\left[a\,s^{\sigma}\left|\begin{array}{l l}
    (a_p,A_p) \\
    (b_q,B_q)
  \end{array}\right.\right]\right](t)=t^{\rho-1}\,H_{p+1,q}^{m,n}\left[\frac{a}{t^{\sigma}}\left|\begin{array}{l l}
    (a_p,A_p), (\rho,\sigma) \\
    (b_q,B_q)
  \end{array}\right.\right].
\end{eqnarray}
The Fox $H$-functions have the following property~\cite{saxena_book}
\begin{eqnarray}\label{h pr1}
    z^{k}\,H_{p,q}^{m,n}\left[z\left|\begin{array}{l l}
    (a_p,A_p)\\
    (b_q,B_q)
  \end{array}\right.\right]
    =H_{p,q}^{m,n}\left[z\left|\begin{array}{l l}
    (a_p+k A_p,A_p)\\
    (b_q+k B_q,B_q)
  \end{array}\right.\right].
\end{eqnarray}






\begin{thebibliography}{10}

\bibitem{stojkoski2019cooperation}
Viktor Stojkoski, Zoran Utkovski, Lasko Basnarkov, and Ljupco Kocarev.
\newblock Cooperation dynamics in networked geometric brownian motion.
\newblock {\em Physical Review E}, 99(6):062312, 2019.

\bibitem{stojkoski2019evolution}
Viktor Stojkoski, Marko Karbevski, Zoran Utkovski, Lasko Basnarkov, and Ljupco
  Kocarev.
\newblock Evolution of cooperation in populations with heterogeneous
  multiplicative resource dynamics.
\newblock {\em arXiv preprint arXiv:1912.09205}, 2019.

\bibitem{peters2013ergodicity}
Ole Peters and William Klein.
\newblock Ergodicity breaking in geometric brownian motion.
\newblock {\em Physical review letters}, 110(10):100603, 2013.

\bibitem{aitchison1957lognormal}
John Aitchison and James~AC Brown.
\newblock {\em The lognormal distribution with special reference to its uses in
  economics}.
\newblock Cambridge Univ. Press, 1957.

\bibitem{redner1990random}
Sidney Redner.
\newblock Random multiplicative processes: An elementary tutorial.
\newblock {\em American Journal of Physics}, 58(3):267--273, 1990.

\bibitem{black1973pricing}
Fischer Black and Myron Scholes.
\newblock The pricing of options and corporate liabilities.
\newblock {\em Journal of Political Economy}, 81(3):637--654, 1973.

\bibitem{merton1975optimum}
Robert~C Merton.
\newblock Optimum consumption and portfolio rules in a continuous-time model.
\newblock In {\em Stochastic Optimization Models in Finance}, pages 621--661.
  Elsevier, 1975.

\bibitem{merton1976option}
Robert~C Merton.
\newblock Option pricing when underlying stock returns are discontinuous.
\newblock {\em Journal of Financial Economics}, 3(1-2):125--144, 1976.

\bibitem{1}
Ole Peters.
\newblock Optimal leverage from non-ergodicity.
\newblock {\em Quantitative Finance}, 11(11):1593--1602, 2011.

\bibitem{oshanin2012two}
Gleb Oshanin and Gregory Schehr.
\newblock Two stock options at the races: Black--scholes forecasts.
\newblock {\em Quantitative Finance}, 12(9):1325--1333, 2012.

\bibitem{procaccia}
HGE Hentschel and Itamar Procaccia.
\newblock Fractal nature of turbulence as manifested in turbulent diffusion.
\newblock {\em Physical Review A}, 27(2):1266, 1983.

\bibitem{7}
M~Sc~Mario Heidern{\"a}tsch.
\newblock {\em On the diffusion in inhomogeneous systems}.
\newblock PhD thesis, Technischen Universit{\"a}t Chemnitz, 2015.

\bibitem{iomin_prl}
E~Baskin and A~Iomin.
\newblock Superdiffusion on a comb structure.
\newblock {\em Physical Review Letters}, 93(12):120603, 2004.

\bibitem{taleb2007black}
Nassim~Nicholas Taleb.
\newblock {\em The Black Swan: The impact of the highly improbable}, volume~2.
\newblock Random House, 2007.

\bibitem{cox1975notes}
John Cox.
\newblock Notes on option pricing i: Constant elasticity of variance
  diffusions.
\newblock {\em Unpublished note, Stanford University, Graduate School of
  Business}, 1975.

\bibitem{heston1993closed}
Steven~L Heston.
\newblock A closed-form solution for options with stochastic volatility with
  applications to bond and currency options.
\newblock {\em The review of financial studies}, 6(2):327--343, 1993.

\bibitem{hagan2002managing}
Patrick~S Hagan, Deep Kumar, Andrew~S Lesniewski, and Diana~E Woodward.
\newblock Managing smile risk.
\newblock {\em The Best of Wilmott}, 1:249--296, 2002.

\bibitem{matacz2000financial}
Andrew Matacz.
\newblock Financial modeling and option theory with the truncated l{\'e}vy
  process.
\newblock {\em International Journal of Theoretical and Applied Finance},
  3(01):143--160, 2000.

\bibitem{borland2002theory}
Lisa Borland.
\newblock A theory of non-gaussian option pricing.
\newblock {\em Quantitative Finance}, 2(6):415--431, 2002.

\bibitem{borland2004non}
Lisa Borland and Jean-Philippe Bouchaud.
\newblock A non-gaussian option pricing model with skew.
\newblock {\em Quantitative Finance}, 4(5):499--514, 2004.

\bibitem{moriconi2007delta}
L~Moriconi.
\newblock Delta hedged option valuation with underlying non-gaussian returns.
\newblock {\em Physica A: Statistical Mechanics and its Applications},
  380:343--350, 2007.

\bibitem{cassidy2010pricing}
Daniel~T Cassidy, Michael~J Hamp, and Rachid Ouyed.
\newblock Pricing european options with a log student’s t-distribution: A
  gosset formula.
\newblock {\em Physica A: Statistical Mechanics and its Applications},
  389(24):5736--5748, 2010.

\bibitem{basnarkov2018option}
Lasko Basnarkov, Viktor Stojkoski, Zoran Utkovski, and Ljupco Kocarev.
\newblock Option pricing with heavy-tailed distributions of logarithmic
  returns.
\newblock {\em arXiv preprint arXiv:1807.01756}, 2018.

\bibitem{13}
Marcin Magdziarz.
\newblock Black-scholes formula in subdiffusive regime.
\newblock {\em Journal of Statistical Physics}, 136(3):553--564, 2009.

\bibitem{14}
CN~Angstmann, BI~Henry, and AV~McGann.
\newblock Time-fractional geometric brownian motion from continuous time random
  walks.
\newblock {\em Physica A}, 526:121002, 2019.

\bibitem{krzyzanowski2020weighted}
Grzegorz Krzy{\.z}anowski, Marcin Magdziarz, and {\L}ukasz P{\l}ociniczak.
\newblock A weighted finite difference method for subdiffusive black--scholes
  model.
\newblock {\em Computers \& Mathematics with Applications}, 80(5):653--670,
  2020.

\bibitem{scalas2000fractional}
Enrico Scalas, Rudolf Gorenflo, and Francesco Mainardi.
\newblock Fractional calculus and continuous-time finance.
\newblock {\em Physica A: Statistical Mechanics and its Applications},
  284(1-4):376--384, 2000.

\bibitem{raberto2002waiting}
Marco Raberto, Enrico Scalas, and Francesco Mainardi.
\newblock Waiting-times and returns in high-frequency financial data: an
  empirical study.
\newblock {\em Physica A: Statistical Mechanics and its Applications},
  314(1-4):749--755, 2002.

\bibitem{2}
Ole Peters and William Klein.
\newblock Ergodicity breaking in geometric brownian motion.
\newblock {\em Physical Review Letters}, 110(10):100603, 2013.

\bibitem{3}
Viktor Stojkoski, Zoran Utkovski, Lasko Basnarkov, and Ljupco Kocarev.
\newblock Cooperation dynamics in networked geometric brownian motion.
\newblock {\em Physical Review E}, 99(6):062312, 2019.

\bibitem{4}
Jun Wang, Jin-Rong Liang, Long-Jin Lv, Wei-Yuan Qiu, and Fu-Yao Ren.
\newblock Continuous time black--scholes equation with transaction costs in
  subdiffusive fractional brownian motion regime.
\newblock {\em Physica A}, 391(3):750--759, 2012.

\bibitem{5}
Gulnur Karipova and Marcin Magdziarz.
\newblock Pricing of basket options in subdiffusive fractional black--scholes
  model.
\newblock {\em Chaos, Solitons \& Fractals}, 102:245--253, 2017.

\bibitem{aw}
Janusz Gajda and Agnieszka Wy{\l}oma{\'n}ska.
\newblock Geometric brownian motion with tempered stable waiting times.
\newblock {\em Journal of Statistical Physics}, 148(2):296--305, 2012.

\bibitem{6}
N~Leibovich and E~Barkai.
\newblock Infinite ergodic theory for heterogeneous diffusion processes.
\newblock {\em Physical Review E}, 99(4):042138, 2019.

\bibitem{merton1973theory}
Robert~C Merton.
\newblock Theory of rational option pricing.
\newblock {\em The Bell Journal of economics and management science}, pages
  141--183, 1973.

\bibitem{hull2003options}
John~C Hull.
\newblock {\em Options futures and other derivatives}.
\newblock Pearson Education India, 2003.

\bibitem{10}
Steven~G Kou.
\newblock A jump-diffusion model for option pricing.
\newblock {\em Management Science}, 48(8):1086--1101, 2002.

\bibitem{11}
John Cox.
\newblock Notes on option pricing i: Constant elasticity of variance
  diffusions.
\newblock {\em Unpublished note, Stanford University, Graduate School of
  Business}, 1975.

\bibitem{12}
Steven~L Heston.
\newblock A closed-form solution for options with stochastic volatility with
  applications to bond and currency options.
\newblock {\em The Review of Financial Studies}, 6(2):327--343, 1993.

\bibitem{20}
Francesco Mainardi.
\newblock {\em Fractional calculus and waves in linear viscoelasticity: an
  introduction to mathematical models}.
\newblock World Scientific, 2010.

\bibitem{fogedby1994langevin}
Hans~C Fogedby.
\newblock Langevin equations for continuous time l{\'e}vy flights.
\newblock {\em Physical Review E}, 50(2):1657, 1994.

\bibitem{li2016option}
Chao Li.
\newblock {\em Option pricing with generalized continuous time random walk
  models}.
\newblock PhD thesis, Queen Mary University of London, 2016.

\bibitem{feller2008introduction}
Willliam Feller.
\newblock {\em An introduction to probability theory and its applications, vol
  2}.
\newblock John Wiley \& Sons, 2008.

\bibitem{magdziarz2007fractional}
Marcin Magdziarz, Aleksander Weron, and Karina Weron.
\newblock Fractional fokker-planck dynamics: Stochastic representation and
  computer simulation.
\newblock {\em Physical Review E}, 75(1):016708, 2007.

\bibitem{magdziarz2008equivalence}
Marcin Magdziarz, Aleksander Weron, and Joseph Klafter.
\newblock Equivalence of the fractional fokker-planck and subordinated langevin
  equations: the case of a time-dependent force.
\newblock {\em Physical review letters}, 101(21):210601, 2008.

\bibitem{garra2018prabhakar}
Roberto Garra and Roberto Garrappa.
\newblock The prabhakar or three parameter mittag--leffler function: Theory and
  application.
\newblock {\em Communications in Nonlinear Science and Numerical Simulation},
  56:314--329, 2018.

\bibitem{sornette1996stock}
Didier Sornette, Anders Johansen, and Jean-Philippe Bouchaud.
\newblock Stock market crashes, precursors and replicas.
\newblock {\em Journal de Physique I}, 6(1):167--175, 1996.

\bibitem{mura2008non}
Antonio Mura, Murad~S Taqqu, and Francesco Mainardi.
\newblock Non-markovian diffusion equations and processes: analysis and
  simulations.
\newblock {\em Physica A: Statistical Mechanics and its Applications},
  387(21):5033--5064, 2008.

\bibitem{mura2008characterizations}
Antonio Mura and Gianni Pagnini.
\newblock Characterizations and simulations of a class of stochastic processes
  to model anomalous diffusion.
\newblock {\em Journal of Physics A: Mathematical and Theoretical},
  41(28):285003, 2008.

\bibitem{sposini2018random}
Vittoria Sposini, Aleksei~V Chechkin, Flavio Seno, Gianni Pagnini, and Ralf
  Metzler.
\newblock Random diffusivity from stochastic equations: comparison of two
  models for brownian yet non-gaussian diffusion.
\newblock {\em New Journal of Physics}, 20(4):043044, 2018.

\bibitem{15}
Marcin Magdziarz and Janusz Gajda.
\newblock Anomalous dynamics of black--scholes model time-changed by inverse
  subordinators.
\newblock {\em Acta Physica Polonica B}, 43(5), 2012.

\bibitem{16}
Ralf Metzler and Joseph Klafter.
\newblock The random walk's guide to anomalous diffusion: a fractional dynamics
  approach.
\newblock {\em Physics Reports}, 339(1):1--77, 2000.

\bibitem{17}
E~Barkai.
\newblock Fractional fokker-planck equation, solution, and application.
\newblock {\em Physical Review E}, 63(4):046118, 2001.

\bibitem{18}
Mark~M Meerschaert, David~A Benson, Hans-Peter Scheffler, and Boris Baeumer.
\newblock Stochastic solution of space-time fractional diffusion equations.
\newblock {\em Physical Review E}, 65(4):041103, 2002.

\bibitem{schulz2014aging}
Johannes~HP Schulz, Eli Barkai, and Ralf Metzler.
\newblock Aging renewal theory and application to random walks.
\newblock {\em Physical Review X}, 4(1):011028, 2014.

\bibitem{schilling2012bernstein}
Ren{\'e}~L Schilling, Renming Song, and Zoran Vondracek.
\newblock {\em Bernstein functions: theory and applications}, volume~37.
\newblock Walter de Gruyter, 2012.

\bibitem{19}
Trifce Sandev, Ralf Metzler, and Aleksei Chechkin.
\newblock From continuous time random walks to the generalized diffusion
  equation.
\newblock {\em Fractional Calculus and Applied Analysis}, 21(1):10--28, 2018.

\bibitem{sandev2017beyond}
Trifce Sandev, Igor~M Sokolov, Ralf Metzler, and Aleksei Chechkin.
\newblock Beyond monofractional kinetics.
\newblock {\em Chaos, Solitons \& Fractals}, 102:210--217, 2017.

\bibitem{prabhakar}
Tilak~Raj Prabhakar.
\newblock A singular integral equation with a generalized mittag leffler
  function in the kernel.
\newblock {\em Yokohama Mathematical Journal}, 19:7--15, 1971.

\bibitem{sandev2015distributed}
Trifce Sandev, Aleksei~V Chechkin, Nickolay Korabel, Holger Kantz, Igor~M
  Sokolov, and Ralf Metzler.
\newblock Distributed-order diffusion equations and multifractality: Models and
  solutions.
\newblock {\em Physical Review E}, 92(4):042117, 2015.

\bibitem{cherstvy}
Andrey~G Cherstvy, Deepak Vinod, Erez Aghion, Aleksei~V Chechkin, and Ralf
  Metzler.
\newblock Time averaging, ageing and delay analysis of financial time series.
\newblock {\em New Journal of Physics}, 19(6):063045, 2017.

\bibitem{el2018quantitative}
Omar El~Euch.
\newblock {\em Quantitative Finance under rough volatility}.
\newblock PhD thesis, Sorbonne universit{\'e}, 2018.

\bibitem{dd1}
Rohit Jain and Kizhakeyil~L Sebastian.
\newblock Diffusion in a crowded, rearranging environment.
\newblock {\em The Journal of Physical Chemistry B}, 120(16):3988--3992, 2016.

\bibitem{dd2}
Aleksei~V Chechkin, Flavio Seno, Ralf Metzler, and Igor~M Sokolov.
\newblock Brownian yet non-gaussian diffusion: from superstatistics to
  subordination of diffusing diffusivities.
\newblock {\em Physical Review X}, 7(2):021002, 2017.

\bibitem{dd3}
Vittoria Sposini, Aleksei~V Chechkin, Flavio Seno, Gianni Pagnini, and Ralf
  Metzler.
\newblock Random diffusivity from stochastic equations: comparison of two
  models for brownian yet non-gaussian diffusion.
\newblock {\em New Journal of Physics}, 20(4):043044, 2018.

\bibitem{dd4}
Wei Wang, Andrey~G Cherstvy, Xianbin Liu, and Ralf Metzler.
\newblock Anomalous diffusion and nonergodicity for heterogeneous diffusion
  processes with fractional gaussian noise.
\newblock {\em Physical Review E}, 102(1):012146, 2020.

\bibitem{dd5}
Wei Wang, Flavio Seno, Igor~M Sokolov, Aleksei~V Chechkin, and Ralf Metzler.
\newblock Unexpected crossovers in correlated random-diffusivity processes.
\newblock {\em New Journal of Physics}, 22(8):083041, 2020.

\bibitem{submitted}
Trifce Sandev, Alexander Iomin, and Kocarev Ljupco.
\newblock Hitting times in turbulent diffusion due to multiplicative noise.
\newblock {\em Physical Review E}, 102(4):042109, 2020.

\bibitem{8}
Fritz Oberhettinger.
\newblock {\em Tables of Mellin transforms}.
\newblock Springer Science \& Business Media, 2012.

\bibitem{18_2}
Christopher~N Angstmann, Isaac~C Donnelly, Bruce~Ian Henry, TAM Langlands, and
  Peter Straka.
\newblock Generalized continuous time random walks, master equations, and
  fractional fokker--planck equations.
\newblock {\em SIAM Journal on Applied Mathematics}, 75(4):1445--1468, 2015.

\bibitem{18_3}
Christopher~N Angstmann, Isaac~C Donnelly, and Bruce~I Henry.
\newblock Continuous time random walks with reactions forcing and trapping.
\newblock {\em Mathematical Modelling of Natural Phenomena}, 8(2):17--27, 2013.

\bibitem{19_2}
Trifce Sandev, Aleksei Chechkin, Holger Kantz, and Ralf Metzler.
\newblock Diffusion and fokker-planck-smoluchowski equations with generalized
  memory kernel.
\newblock {\em Fractional Calculus and Applied Analysis}, 18(4):1006, 2015.

\bibitem{saxena_book}
Arakaparampil~M Mathai, Ram~Kishore Saxena, and Hans~J Haubold.
\newblock {\em The H-function: theory and applications}.
\newblock Springer Science \& Business Media, 2009.

\end{thebibliography}




\end{document}